\documentclass[preprint,5p,twocolumn,10pt]{elsarticle}

\usepackage{xcolor}
\usepackage[pdftex,hypertexnames=false,linktocpage=true]{hyperref}
\hypersetup{pdfauthor={Satvir Kaur, Sreeraj Nair, Chandan Mondal, Jiangshan Lan, Xingbo Zhao, J. P. B. C. de Melo, Tobias Frederico},colorlinks=true,linkcolor=red,anchorcolor=darkgreen,citecolor=green!50!blue,filecolor=darkgreen,urlcolor=green,bookmarksnumbered=true,pdfview=FitB}
\usepackage{lipsum}
\usepackage{mathtools,cuted}
\usepackage{cuted}
\usepackage[normalem]{ulem}

\usepackage{graphicx}
\usepackage{float}
\usepackage{subfigure}

\usepackage{array}
\usepackage{booktabs} 
\usepackage{arydshln}
\usepackage{multirow}
\usepackage{makecell} 

\def\arraystretch{1.8}

\def\psla{\slash \! \!\! }

\newcommand{\be}{\begin{equation}}  
\newcommand{\ee}{\end{equation}}  

\newcommand{\ba}{\begin{align}}  
\newcommand{\ea}{\end{align}}

\hypersetup{colorlinks=true, citecolor=blue, urlcolor=blue, linkcolor=blue}

\usepackage{braket}

\usepackage{comment}
\begin{document}

\begin{frontmatter}
\title{Baryon Bethe–Salpeter Equation in Minkowski-Space QCD$_2$}

\author[inst1,inst2]{Satvir Kaur}
\ead{satvir@impcas.ac.cn}

\author[inst1,inst2]{Sreeraj Nair}
\ead{sreeraj@impcas.ac.cn}

\author[inst1,inst2]{Chandan Mondal}
\ead{mondal@impcas.ac.cn}

\author[inst1,inst2,inst3]{Jiangshan Lan}
\ead{jiangshanlan@impcas.ac.cn}

\author[inst1,inst2,inst3]{Xingbo Zhao}
\ead{xbzhao@impcas.ac.cn}

\author[inst5,inst6]{J. P. B. C. de Melo}
\ead{joao.mello@cruzeirodosul.edu.br}

\author[inst6]{Tobias~Frederico}
\ead{tobias@ita.br}

\address[inst1]{State Key Laboratory of Heavy Ion Science and Technology, Institute of Modern Physics, Chinese Academy of Sciences, Lanzhou 730000, China}
\address[inst2]{School of Nuclear Science and Technology, University of Chinese Academy of Sciences, Beijing 100049, China}
\address[inst3]{Advanced Energy Science and Technology Guangdong Laboratory, Huizhou, Guangdong 516000, China}
\address[inst5]{Laborat\'orio de F\'isica Te\'orica e Computacional--LFTC, Universidade Cruzeiro do Sul / Universidade Cidade de S\~ao Paulo, 015060-000, S\~ao Paulo, SP, Brazil}
\address[inst6]{Instituto Tecnol\'ogico de Aeron\'autica, 12.228-900, S\~ao Jos\'e dos Campos, SP, Brazil}

\begin{abstract}
	We study the three-quark ladder Bethe--Salpeter equation in Minkowski-space QCD$_2$ in the light-cone gauge. Using the quasi-potential expansion, we project the baryon equation onto the light front and show that, at leading order in the valence truncation, the resulting mass-squared eigenvalue equation is equivalent to the Bars--Durgut equation. We also derive the endpoint power-law behavior of the valence wave function in terms of the quark mass and coupling, closely paralleling the original 't Hooft analysis for mesons. The resulting three-quark equation is solved numerically for $N_c=3$, and the ground-state baryon mass is found to be in reasonable agreement with previous light-cone quantization results in QCD$_2$, suggesting that the valence sector provides the dominant contribution to the ground state. The excited-state spectrum further yields a Regge trajectory that captures the overall trend of the experimental nucleon spectrum, and we compute selected structure observables, including parton distribution functions, double distribution amplitudes, and coordinate-space densities. This framework provides a useful confining test bed for Minkowski-space bound-state methods and for future developments toward confining formulations in (3+1) dimensions beyond the valence truncation.
\end{abstract}
\end{frontmatter}

\section{Introduction}
The 't Hooft equation for mesons~\cite{tHooft:1974pnl}, derived in large-$N_c$ QCD in (1+1) dimensions (QCD$_2$), provides a clear example of confinement in a relativistic quantum field theory. Beyond offering an analytically tractable bound-state equation, it has influenced numerous developments in hadron structure and light-front dynamics. A key extension was the light-cone quantization study of QCD$_2$ by Hornbostel, Brodsky, and Pauli~\cite{Hornbostel:1988fb}, who solved the theory for finite $N_c$ and obtained meson and baryon spectra. This work laid the foundation for the Basis Light-Front Quantization (BLFQ) approach, which represents hadronic states directly in light-front (LF) Fock space and treats them nonperturbatively~\cite{Vary:2009gt,Vary:2025yqo,Mondal:2019jdg,Xu:2021wwj,Xu:2022yxb,Xu:2024sjt,Lan:2019vui,Lan:2021wok}.

Alongside LF Hamiltonian methods, the Bethe--Salpeter equation (BSE) provides a covariant framework for relativistic bound states in Minkowski space. Significant progress has been made in solving BSEs with nonconfining kernels using approaches such as the Nakanishi integral representation~\cite{Nakarev,Kusaka:1995za,Karmanov:2005nv,Frederico:2011ws} and direct treatments of singularities and cuts~\cite{Carbonell:2014dwa,Eichmann:2021vnj}, see reviews in Ref.~\cite{dePaula:2026gtb}. These formulations resum infinitely many LF time-ordered intermediate states for ladder or cross-ladder kernels. Extending them to genuinely confining theories, however, remains a major challenge.

In this context, QCD$_2$ provides a valuable laboratory, with confinement realized explicitly in both the large-$N_c$ meson equation~\cite{tHooft:1974pnl} and light-cone quantized QCD$_2$ at finite $N_c$~\cite{Hornbostel:1988fb}. This makes it an ideal setting to study the LF reduction of a confining Minkowski-space Bethe--Salpeter framework while preserving the essential spectral and structural features of the bound-state problem.

The influence of the 't Hooft equation extends beyond two-dimensional QCD. It has recently played a key role in modeling longitudinal dynamics within holographic and LF approaches to hadrons~\cite{Li:2021jqb,Li2022,deTeramond:2021yyi,Lyubovitskij:2022rod,Weller:2021wog,Ahmady:2021lsh,Rinaldi:2022dyh,Gurjar:2025kcp} and nuclei~\cite{Kaur:2025css}. In particular, incorporating longitudinal confinement and dynamical chiral symmetry breaking has enabled successful descriptions of meson spectroscopy and structure, including light and heavy mesons and gluonic observables in the pion~\cite{Kaur:2025gyr,Gurjar:2024wpq,Ahmady:2022dfv,Ahmady:2021yzh, Gurjar:2025kcp}, underscoring the importance of longitudinal dynamics in LF bound-state formulations.

For baryons, the situation is less developed. Unlike the large-$N_c$ meson case, the baryon problem is not closed in the valence sector and generally requires coupling to higher Fock components. Nevertheless, an effective three-quark equation in QCD$ _2$ was derived by Bars~\cite{Bars:1976re,Bars:1976nk} and Durgut~\cite{Durgut:1976bc}, using string-inspired methods and planar Feynman diagrams, respectively. Webber~\cite{Webber:1979na} later solved the Bars--Durgut equation (BDE) numerically for vanishing quark mass, obtaining a Regge-like pattern for excited baryons, though the ground-state mass vanished. These results highlight the relevance of the BDE and motivate its derivation from a Minkowski-space three-quark Bethe--Salpeter framework.

In this work, we investigate the three-quark BSE in Minkowski-space QCD$_2$ with massless gluon exchange in the light-cone gauge. The central novelty of our formulation is the embedding of the classic BDE within the covariant Minkowski-space ladder BSE framework and its derivation through a systematic LF reduction based on the quasi-potential (QP) expansion~\cite{Sales:1999ec,Sales:2001gk,Marinho:2007zz,Marinho:2008zza,Guimaraes:2014kor,Ydrefors:2021mky}. This approach provides a controlled connection between the full relativistic three-body dynamics and the LF valence sector, revealing the BDE as the leading-order (LO) approximation to an effective mass-squared operator acting on the valence wave function. Beyond recovering the BDE, the QP expansion establishes a systematic and improvable framework for incorporating higher-Fock-state contributions order by order, thereby extending the original Bars--Durgut formulation.

We also analyze the endpoint behavior of the valence wave function as a function of the quark mass and coupling strength, and introduce a constituent quark mass to facilitate numerical solutions. The formalism is then applied to calculate the ground- and excited-state baryon spectrum, Regge trajectories, valence parton distribution functions (PDFs) evolved to experimental scales, double distribution amplitudes (DDAs), and coordinate-space densities.

\section{Baryon BSE in QCD$_{2}$}\label{sec:BSE}
In QCD$_2$, the three-quark BSE in the ladder approximation can be written as
\begin{multline}\label{eq:bse_1}
		\Psi_M(k_1,k_2,k_3)= -\frac{i\,g_{\rm H}^2}{(2\pi)^2}\,
	S_1(k_1)\otimes S_2(k_2)\otimes S_3(k_3) \\
	\otimes \sum_{ijk}\sum_a S_i^{-1}(k_i)\otimes
	\int d^2k'_j\,\\
	{}\times
	\frac{(\gamma_j^+T_j^a)\otimes(\gamma_k^+T_k^a)}
	{(k_j^+-k_j^{\prime +})^2}
	\,\Psi_M(k_i,k'_j,k'_k)\,,
\end{multline}
where the first sum runs over the cyclic permutations of
$\{i,j,k\}$, while the adjoint color index $a$ is summed over. We denote the gauge coupling prior to color projection by $g_{\rm H}$ to distinguish it from the effective pair-kernel coupling introduced below. The quark--gluon vertex is therefore $ig_{\rm H}\gamma^+T^a$. In Ref.~\cite{Hornbostel:1988fb}, the same fundamental coupling is denoted by $g$. The quark propagator is
\begin{equation}
	S(k)=i\,\frac{\psla k_{on}+m}{k^+\left(k^--k^-_{on}+\frac{i\varepsilon}{k^+}\right)}
	+i\,\frac{\gamma^+}{2k^+}\,,
\end{equation}
	with the second term corresponding to the instantaneous part of the LF propagator.

The total momentum of the baryon satisfies
\begin{equation}
	K=k_1+k_2+k_3=k_i+k'_j+k'_k\,,
\end{equation}
with invariant mass
$	K^2=M^2$,
where $M$ denotes the baryon mass. As in the meson case studied by 't Hooft~\cite{tHooft:1974pnl}, quark self-energy contributions are needed to regularize endpoint divergences. We include these terms later.

\begin{figure}[htp]
	\centering
	\includegraphics[width=\columnwidth]{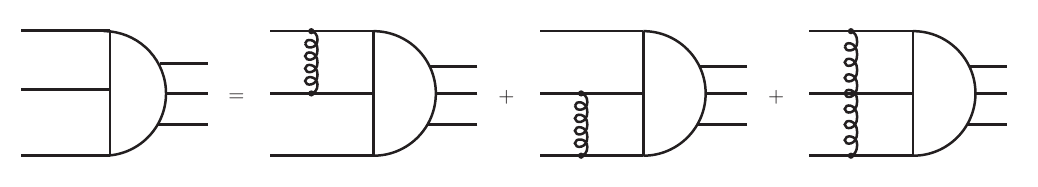} 
	\caption{Diagrammatic representation of the three-quark BSE in QCD$_2$ in the light-cone gauge. The full baryon amplitude, shown on the left, is written as the sum of the three pairwise instantaneous gluon-exchange contributions between quark pairs.}
	\label{fig:BSE}
\end{figure}

A diagrammatic representation of Eq.~\eqref{eq:bse_1} is shown in Fig.~\ref{fig:BSE}, where the vertical gluon lines denote instantaneous exchanges in the light-cone gauge. It is important to emphasize that Eq.~\eqref{eq:bse_1} is not identical to the BDE. The present Bethe--Salpeter framework contains, in principle, intermediate states beyond the valence sector, whereas the BDE corresponds to the valence three-quark dynamics. This distinction becomes explicit after the LF projection presented in the next section, which is detailed in \ref{sec:LFBSE} using the QP expansion (for a review, see~\cite{dePaula:2026gtb}).

\section{Leading order baryon equation}\label{sec:LOBaryoneq}
The QP expansion of the LF-projected BSE yields an effective squared-mass operator acting on the valence component of the wave function by eliminating the relative LF-time of the propagating particles in Minkowski space. The resulting effective interaction systematically couples the valence state to infinitely many higher Fock sectors, which can be computed recursively within the QP expansion. This approach is closely related to the Iterated Resolvent Method~\cite{Brodsky:1997de}, reducing the LF Hamiltonian diagonalization to the valence sector. Technical details of the LF projection and the derivation of the leading-order effective interaction via truncation of the QP expansion are given in~\ref{sec:LFBSE}.

Color indices and generators are suppressed in the reduced valence equation. For a quark pair in a color-singlet SU(3) baryon, the attractive color coefficient is~\cite{Zahed:2026qcd2}
\begin{equation}
	c_{qq}\equiv-\langle \boldsymbol{T}_i\!\cdot\!\boldsymbol{T}_j\rangle
	=\frac{N_c+1}{2N_c}=\frac{2}{3}\,.
\end{equation}
After projection onto the color-singlet baryon, we absorb this coefficient into the effective pair-kernel coupling $g$ used in the reduced equation. The fundamental coupling $g_{\rm H}$ and the color-projected coupling $g$ are related by
\begin{equation}\label{eq:coupling_dictionary}
	g^2=c_{qq}g_{\rm H}^2\,,\qquad
	g=\sqrt{\frac{2}{3}}\,g_{\rm H}\,.
\end{equation}
In the reduced equations below, $g$ always denotes this color-projected pair-kernel coupling.

Considering  the LO effective interaction from Eq.~\eqref{Eq:w_LP4}, the LF projected BSE for the valence three-quark amplitude, Eq.~\eqref{eq:LFprojBS}, takes the form
\begin{align}\label{eq:LOval_1}
		&\chi_3(k_1^+,k_2^+,k_3^+)
		=
		i\frac{g^2}{4\pi}\,
		g_0(k_1^+,k_2^+,k_3^+)\,
		\gamma_1^+\otimes\gamma_2^+\otimes\gamma_3^+
		\nonumber\\ &\times
		\Bigg\{
		\int_0^{K^+-k_3^+} dk_1^{\prime +}\,
		\frac{\chi_3(k_1^{\prime +},k_2^{\prime\prime +},k_3^+) - \chi_3(k_1^+,k_2^+,k_3^+)}
		{(k_1^+-k_1^{\prime +})^2}
		\nonumber\\
		&+\int_0^{K^+-k_1^+} dk_2^{\prime +}\,
		\frac{\chi_3(k_1^+,k_2^{\prime +},k_3^{\prime\prime +}) - \chi_3(k_1^+,k_2^+,k_3^+)}
		{(k_2^+-k_2^{\prime +})^2}
		\\
		&+\int_0^{K^+-k_2^+} dk_3^{\prime +}\,
		\frac{\chi_3(k_1^{\prime\prime +},k_2^+,k_3^{\prime +}) - \chi_3(k_1^+,k_2^+,k_3^+)}
		{(k_3^+-k_3^{\prime +})^2}
		\Bigg\}\, ,\nonumber
	\end{align}
where  the operator $g_0(k_1^+,k_2^+,k_3^+)$   is defined in Eq.~\eqref{propfl3}.
The subtracted terms in Eq.~\eqref{eq:LOval_1} serve to regularize the integrals, directly analogous to the treatment in the 't Hooft equation for meson~\cite{tHooft:1974pnl}. The double-primed longitudinal momenta are determined by the conservation of the total plus momentum.
We now introduce
\begin{equation}
	\psi(k_1^+,k_2^+,k_3^+)
	=	\gamma_1^+\otimes\gamma_2^+\otimes\gamma_3^+\,
	\chi_3(k_1^+,k_2^+,k_3^+)\,,
\end{equation}
and use the identity $\gamma^+(\psla k_{\mathrm{on}}+m)\gamma^+ = 2k^+\gamma^+$.

Furthermore, using the momentum fractions $x_i={k_i^+}/{K^+}$, which satisfy $x_1+x_2+x_3=1$,
the LF equation, equivalent to BDE, can be written as an eigenvalue equation for the baryon mass-squared operator, 
\begin{align}\label{eq:LOval_3}
&M^2\,\psi(x_1,x_2,x_3)
	=
	\left(
	\frac{m^2}{x_1}+\frac{m^2}{x_2}+\frac{m^2}{x_3}
	\right)\psi(x_1,x_2,x_3)
	\nonumber\\
	&-\frac{g^2}{\pi}
	\Bigg\{
	\int_0^{1-x_3} dx_1'\,
	\frac{\psi(x_1',x_2'',x_3)-\psi(x_1,x_2,x_3)}
	{(x_1-x_1')^2}
	\nonumber\\
	&+\int_0^{1-x_1} dx_2'\,
	\frac{\psi(x_1,x_2',x_3'')-\psi(x_1,x_2,x_3)}
	{(x_2-x_2')^2}
	\nonumber\\
	&+\int_0^{1-x_2} dx_3'\,
	\frac{\psi(x_1'',x_2,x_3')-\psi(x_1,x_2,x_3)}
	{(x_3-x_3')^2}
	\Bigg\}\, .
\end{align}
Here the double-primed momentum fractions are determined by longitudinal momentum conservation,
\begin{equation}
\begin{aligned}
	x_2''=1-x_1'-x_3\,,
	&\qquad
	x_3''=1-x_1-x_2'\,,\\
	x_1''=&1-x_2-x_3'\,.
\end{aligned}
\end{equation}
The integrals are understood as principal values.

\begin{figure}[t]
    \centering
\includegraphics[width=0.23\textwidth]{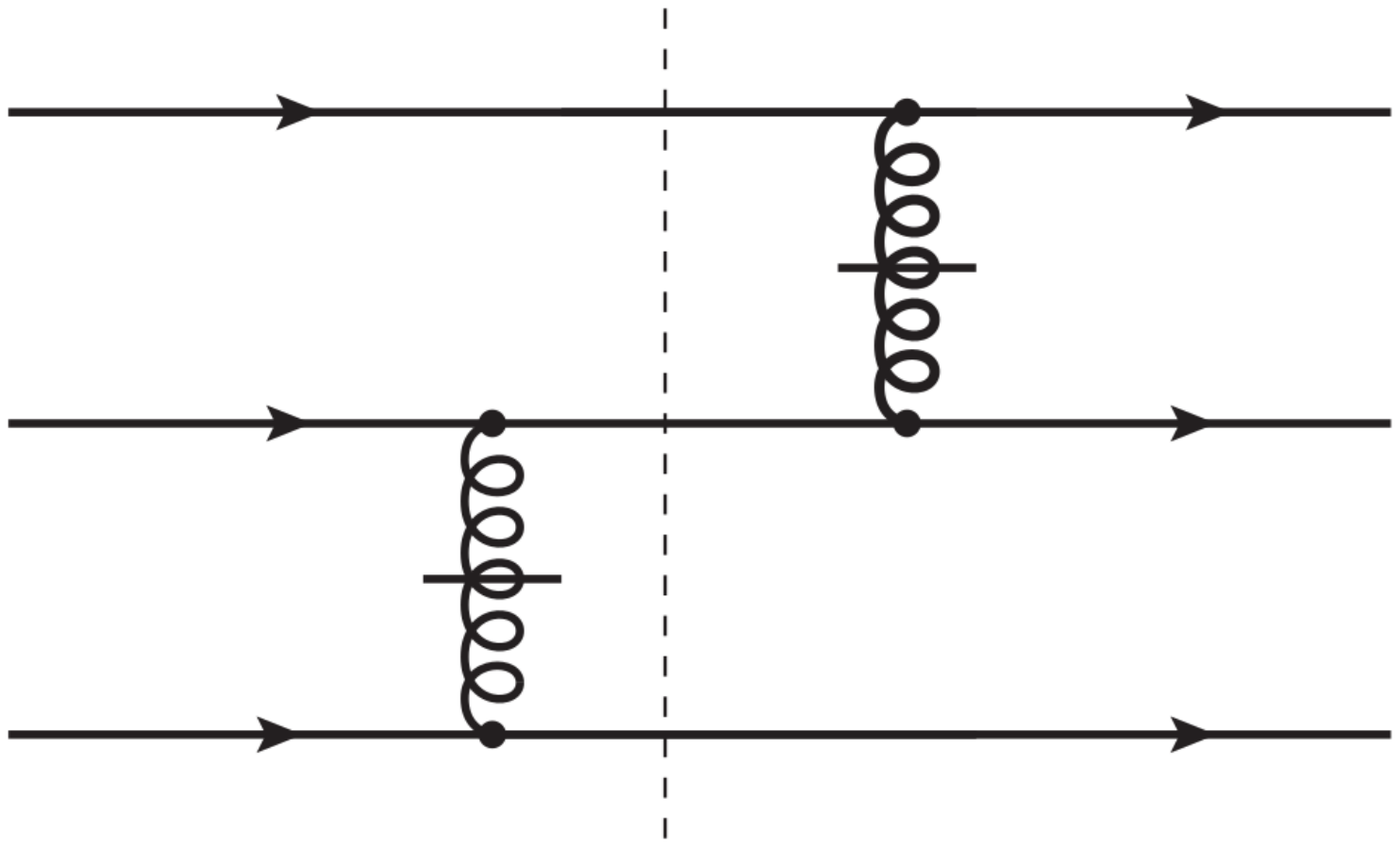}   
\includegraphics[width=0.23\textwidth]{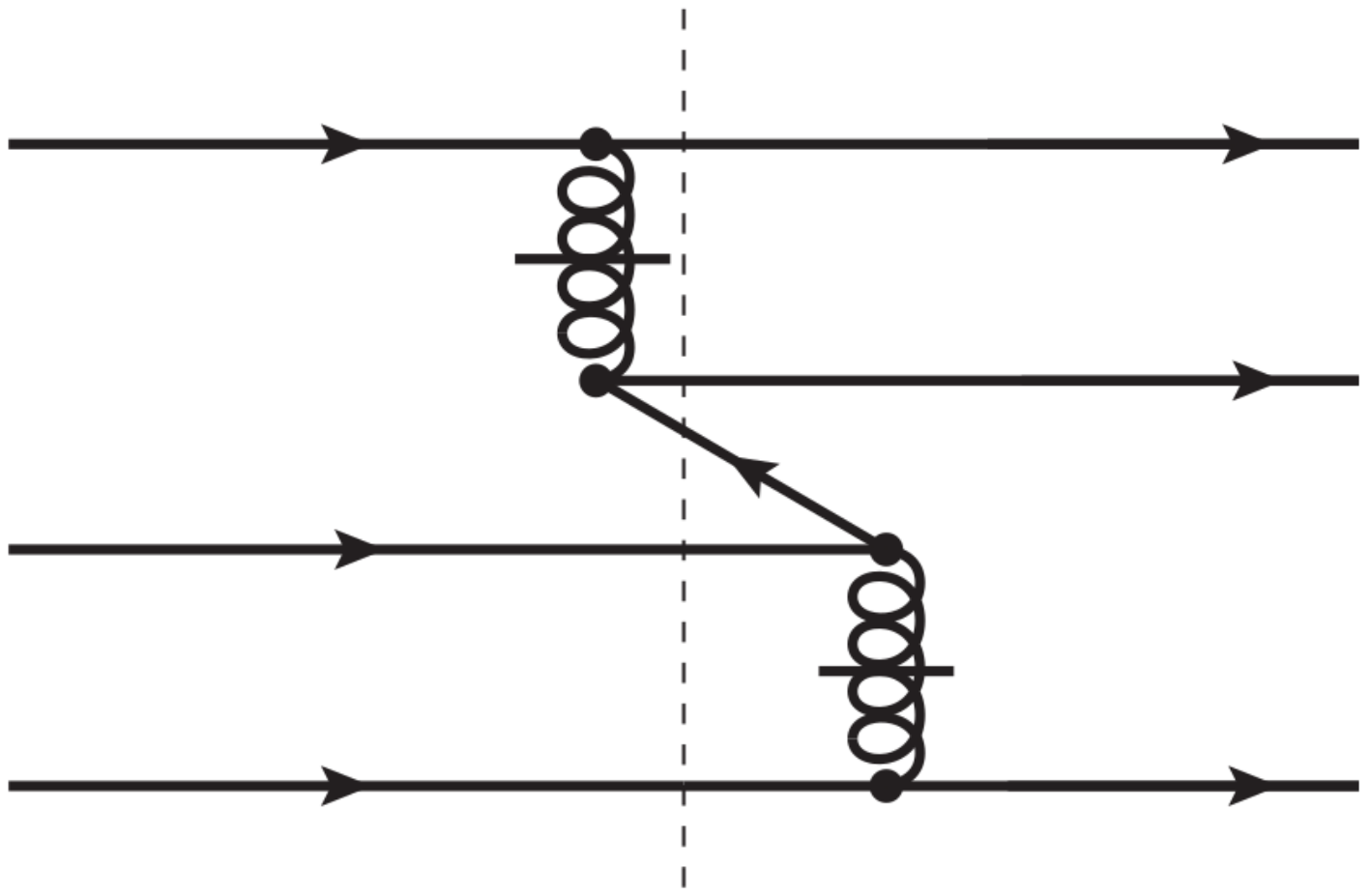}   
    \caption{Schematic iteration of the LO valence equation. Left: an intermediate three-quark LF propagation retained in Eq.~\eqref{eq:LOval_3}. Right: a representative higher-Fock intermediate state containing an additional quark--antiquark pair. It is not included at LO and appears in one of the NLO terms (see Eq.~\eqref{Eq:w_NLO1}).}
    \label{fig:kernel_LO}
\end{figure}

Equation~\eqref{eq:LOval_3} represents the LO valence three-quark equation in QCD$_2$. It includes only intermediate three-particle LF propagation, as illustrated in the left panel of Fig.~\ref{fig:kernel_LO}, while contributions from higher Fock sectors are omitted at this level. An example is shown schematically in the right panel, where the intermediate state contains an additional quark--antiquark pair. Such contributions are not included in Eq.~\eqref{eq:LOval_3}, but they arise naturally beyond LO in the QP expansion as irreducible many-body terms in the effective valence-space Hamiltonian.

\subsection{Endpoint behavior}
\begin{figure}[htp]
    \centering
\includegraphics[width=0.48\textwidth]{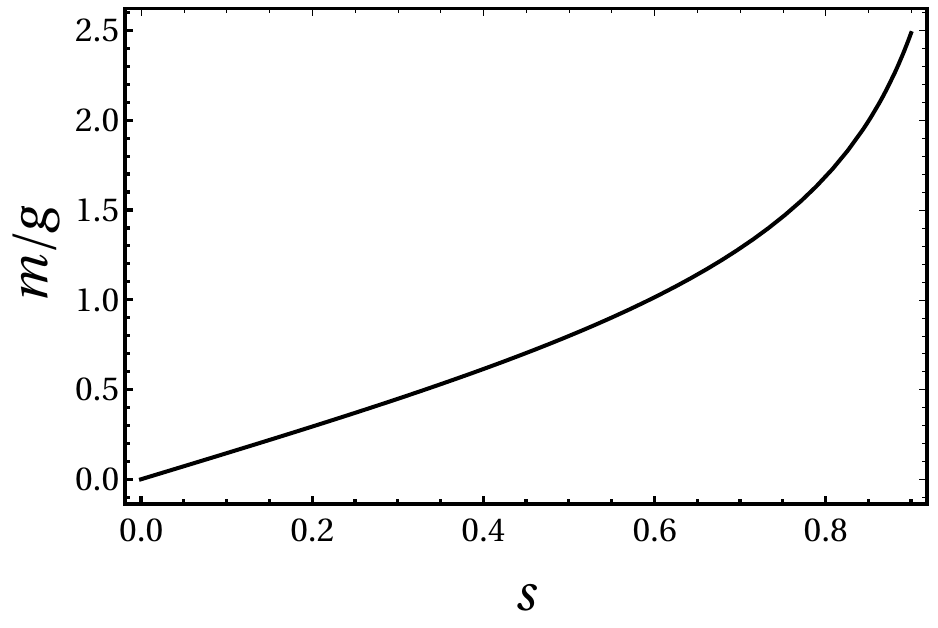}
        \caption{Endpoint exponent $s$ as a function of the ratio $m/g$, obtained from the transcendental Eq.~\eqref{eq:UV}, where  $m$ and $g$ are the mass of quark and the strength of confinement, respectively. }
\label{fig:g_over_m_vs_beta}
\end{figure}

The endpoint behavior of the valence wave function can be derived following the original analysis of the 't Hooft equation~\cite{tHooft:1974pnl}. We begin with the ansatz
\begin{equation}\label{ansatz}
	\psi(x_1,x_2,x_3)=(x_1x_2x_3)^s\,,
\end{equation}
which is expected to describe the asymptotic behavior of Eq.~\eqref{eq:LOval_3} when any one of the momentum fractions tends to zero. It leads to the equation determining the
endpoint exponent,
\begin{equation}\label{eq:UV}
	\frac{m^2}{g^2}\frac{\pi}{2}-1+\pi s\cot(\pi s)=0\,.
\end{equation}
This is the baryon analogue of the 't Hooft transcendental equation. Relative to the meson case, the coefficient of the mass term is modified by a factor of $1/2$. In Fig.~\ref{fig:g_over_m_vs_beta} we display the resulting relation between $m/g$ and $s$ obtained from Eq.~\eqref{eq:UV}.

\section{Numerical method}\label{sec:method}

The LO equation for the valence baryon LF wave function, Eq.~\eqref{eq:LOval_3}, is analogous in form to the 't Hooft equation for mesons. A key difference is that, unlike the meson case in the large-$N_c$ limit, this equation arises from the ladder approximation to the three-body Bethe--Salpeter equation and is therefore not closed. Nevertheless, for phenomenological purposes, we explore Eq.~\eqref{eq:LOval_3} quantitatively. A related numerical approach for LF bound-state problems was introduced in Ref.~\cite{Mo:1992sv} and applied to the massive Schwinger model using the LF Tamm--Dancoff approximation.

To solve Eq.~\eqref{eq:LOval_3}, we expand the eigenstate in a basis of Jacobi polynomials,
\begin{equation}\label{eq:basis_1}
	f_k(\{x_i\})=(x_1x_2x_3)^\beta
	P_k^{\beta,\beta}(x_1)\,
	P_k^{\beta,\beta}(x_2)\,
	P_k^{\beta,\beta}(x_3)\,,
\end{equation}
where $\{x_i\}\equiv\{x_1,x_2,x_3\}$ and $x_3=1-x_1-x_2$. The parameter $\beta$ is later identified with the endpoint exponent $s$ obtained from Eq.~\eqref{eq:UV}. The basis functions are non-orthogonal over the physical integration region, with overlap matrix:
\begin{equation}
	d_{mn}
	=
	\int_0^1 dx_1 \int_0^{1-x_1} dx_2\,
	f_m(\{x_i\})\,f_n(\{x_i\})\,.
\end{equation}
The baryon wave function is expanded as
\begin{equation}\label{eq:basis_2}
	\psi(x_1,x_2,x_3)
	=
	\sum_k a_k\,f_k(x_1,x_2,x_3)\,.
\end{equation}
Since the basis is non-orthogonal, the problem reduces to a generalized eigenvalue problem.

The mass-squared eigenvalue equation is recast as
\begin{equation}\label{eq:H3q}
	H
	=
	T+\frac{g^2}{\pi}\left(V^{(1)}+V^{(2)}+V^{(3)}\right),
\end{equation}
where the matrix elements of the kinetic operator are
\begin{multline}
	T_{mn}
	=
	\int_0^1 dx_1 \int_0^{1-x_1} dx_2
	\left(
	\frac{m^2}{x_1}+\frac{m^2}{x_2}+\frac{m^2}{1-x_1-x_2}
	\right)
	\\ \times
	f_m(x_1,x_2,1-x_1-x_2)\,
	f_n(x_1,x_2,1-x_1-x_2)\,.
\end{multline}
Following the symmetric form of the kernel familiar from the 't Hooft equation~\cite{tHooft:1974pnl}, we write explicitly the interaction matrix element for the quark pair $\{12\}$ as
\begin{align}\label{eq:V1}
	V^{(1)}_{mn} =& \frac{1}{2} \int [{\rm d}x_i] \int_0^{1-x_3}  {\rm d}y_1 	\nonumber\\ &\times\frac{f_m(x_1,x_2,x_3)-f_m(y_1,1-y_1-x_3,x_3)}{x_1-y_1} 	\nonumber\\ &\times \frac{f_n(x_1,x_2,x_3)-f_n(y_1,1-y_1-x_3,x_3)}{x_1-y_1}\,,
\end{align}
where
\begin{equation}
	[dx_i]
	=
	\prod_{i=1}^3 dx_i\,\nonumber
	\delta\!\left(1-\sum_{i=1}^3 x_i\right)
\end{equation}
is the phase-space measure. The matrix elements \(V_{mn}^{(2)}\) and \(V_{mn}^{(3)}\), corresponding to the quark pairs $\lbrace 23 \rbrace$ and $\lbrace 13 \rbrace$, are obtained from Eq.~\eqref{eq:V1} by cyclic permutation of the momentum fractions.

Since the basis functions are totally symmetric, permutation symmetry implies
\begin{equation}
	V_{mn}^{(1)}=V_{mn}^{(2)}=V_{mn}^{(3)}\,.
\end{equation}
It is therefore sufficient in practice to evaluate only \(V_{mn}^{(1)}\), and the Hamiltonian matrix becomes
\begin{equation}
	H_{mn}=T_{mn}+\frac{g^2}{\pi} \sum_{i=1}^3 V_{mn}^{(i)}\,.
\end{equation}
After evaluating the matrix elements, we solve the generalized eigenvalue problem
\begin{equation}
	M^2\sum_n d_{mn}\,a_n
	=
	\sum_n H_{mn}\,a_n\,,
\end{equation}
which yields the mass-squared eigenvalues \(M^2\) and the corresponding baryon eigenstates.

		\begin{figure}[t]
			\centering
	\includegraphics[width=0.48\textwidth]{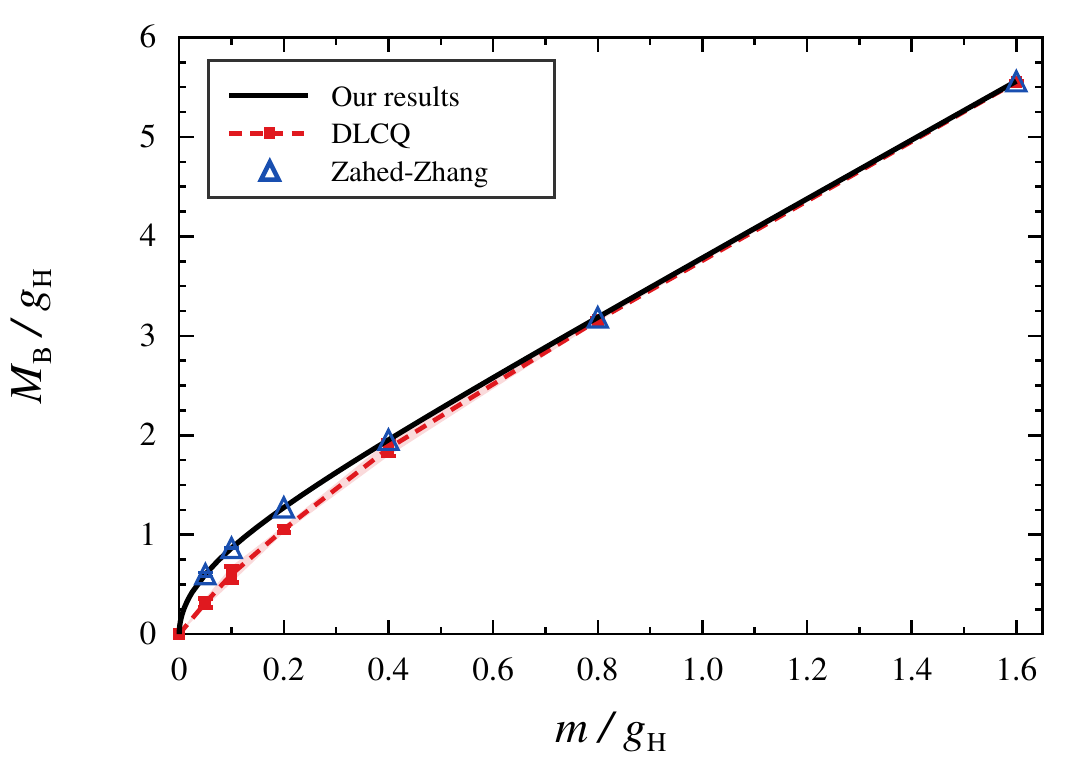}
		\caption{Ground-state baryon mass in SU(3) QCD$_2$ in the common $g_{\rm H}$ convention. The solid black curve is our valence result. The blue triangles show the independent valence calculation of Zahed and Zhang~\cite{Zahed:2026qcd2}. The red squares and shaded band show the continuum values and quoted extrapolation estimates obtained with discretized light-cone quantization (DLCQ) by Hornbostel, Brodsky, and Pauli~\cite{Hornbostel:1988fb}.} 
			\label{fig:MSU3}
		\end{figure}
\section{Baryon phenomenology}\label{sec:Bpheno}
We solve the eigenvalue equation for the three-quark Hamiltonian, Eq.~\eqref{eq:H3q}, using the basis expansions defined in Eqs.~\eqref{eq:basis_1} and \eqref{eq:basis_2}. The calculation is performed with $m = 210~\rm{MeV}$, the effective pair-kernel coupling $g = 330~\rm{MeV}$ defined in Eq.~\eqref{eq:coupling_dictionary}, and optimized $\beta$. The parameters $m$ and $g$ are chosen to reproduce the nucleon mass and the slope of the Regge trajectory.

For Fig.~\ref{fig:MSU3} and
Table~\ref{tab:ground_state_comparison}, we use $D=100$
basis functions and a quadrature order of $Q=320$ to evaluate
the pair-interaction matrix elements. At these values, the
ground-state results are numerically converged and remain
unchanged, within the quoted precision, under further increases
of $D$ and $Q$. The resulting dimensionless ratios are expressed
in the common $g_{\rm H}$ convention using
Eq.~\eqref{eq:coupling_dictionary}. This conversion changes only
the coupling convention used for the comparison; the Hamiltonian
is evaluated with $g=330~\mathrm{MeV}$. The transcendental equation~\eqref{eq:UV} yields $s = 0.413$, in good agreement with the numerical solution, $s \simeq 0.420$.

Figure~\ref{fig:MSU3} compares our valence result with the recent
Faddeev calculation of Zahed and Zhang~\cite{Zahed:2026qcd2} and
the DLCQ calculation of Hornbostel, Brodsky, and
Pauli~\cite{Hornbostel:1988fb}. All three results are expressed
using the same $g_{\rm H}$ convention. Our curve agrees closely
with the Zahed--Zhang values over the displayed range. At every
nonzero tabulated ratio, our values lie above those of Hornbostel
\textit{et al.} At $m/g_{\rm H}=0.4$, $0.8$, and $1.6$, the
relative differences are approximately $4.5\%$, $0.9\%$, and
$0.16\%$, respectively, with the deviation increasing toward the
low-mass region. The numerical comparison is presented in
Table~\ref{tab:ground_state_comparison}.

Since the calculation of Hornbostel \textit{et al.} includes Fock
sectors omitted from the present valence equation, this ordering is
compatible with an attractive net correction to the baryon mass.
The comparison indicates the sign and overall trend of the
omitted-sector effect, while its decomposition into individual
higher-Fock contributions requires an explicit calculation.

\begin{table}[t]
	\centering
	\caption{Ground-state masses $M_{\rm B}/g_{\rm H}$ in the common
coupling convention used in Fig.~\ref{fig:MSU3}. The Zahed--Zhang
values and their quoted uncertainties are taken from Table~III of
Ref.~\cite{Zahed:2026qcd2}.
Following that reference, the results of Hornbostel \textit{et al.}
are converted from the program values $T$ given in Table~I of
Ref.~\cite{Hornbostel:1988fb} using
$M_{\rm B}/g_{\rm H}
=\sqrt{T[(m/g_{\rm H})^2+1/\pi]}$.
The numbers in parentheses in the Hornbostel \textit{et al.}
column denote the propagated uncertainties of the
continuum-extrapolated results. Entries marked by an asterisk
were described in Ref.~\cite{Hornbostel:1988fb} as rough guides.}
	\label{tab:ground_state_comparison}
	\footnotesize
	\renewcommand{\arraystretch}{1.14}
	\setlength{\tabcolsep}{3.0pt}
	\begin{tabular}{cccc}
		\toprule
		$m/g_{\rm H}$ & This work & Zahed--Zhang & Hornbostel \textit{et al.} \\
		\midrule
		0.05 & 0.6018 & 0.603(19) & 0.315(46)$^{\ast}$ \\
		0.10 & 0.8654 & 0.867(3)  & 0.601(82)$^{\ast}$ \\
		0.20 & 1.2733 & 1.274     & 1.054(34) \\
		0.40 & 1.9534 & 1.954     & 1.869(77) \\
		0.80 & 3.1851 & 3.185     & 3.157(15) \\
		1.60 & 5.5610 & 5.560     & 5.552(5) \\
		\bottomrule
	\end{tabular}
\end{table}

Higher-Fock contributions are expected to become more important as $m/g$ decreases. At NLO, the effective valence-space interaction receives contributions mediated by virtual $4q\bar q$ intermediate states,
\begin{equation}
	\delta w^{\mathrm{NLO}}=\sum_{i=1}^{3}\delta w_i^{\mathrm{NLO}}.
\end{equation}
A representative contribution is shown in the right panel of
Fig.~\ref{fig:kernel_LO}. The individual terms, given in Eq.~\eqref{Eq:w_NLO1}, contain contributions such as $V_i\Delta_0V_j$ and $V_i\Delta_0V_k$. For an order-of-magnitude estimate at large $m/g$, we write
\begin{equation}
	\delta M_{\mathrm{B}}^2\sim
	6\frac{g^4}{\pi^2}
	\left\langle
	\frac{x^{-2}_{q_1\bar q}\,x^{-2}_{q_2\bar q}}
	{M_{\mathrm{B}}^2-\sum_{i=1}^5m^2/x_i}
	\right\rangle ,
\end{equation}
Here, the expectation value is evaluated using the LO wave function,
while $x_{q_1\bar q}$ and $x_{q_2\bar q}$ denote the momentum fractions carried by the instantaneous gluon exchanges associated with the creation and annihilation of the virtual pair, respectively; see the right panel of Fig.~\ref{fig:kernel_LO}. At the representative point $m/g\simeq1.6$ and $M_{\mathrm{B}}/g\simeq6$, one has $M_{\mathrm{B}}<5m$, so the intermediate-state denominator is negative. Taking $x_{q_1\bar q}$ and $x_{q_2\bar q}$ to be of order unity gives
\begin{equation}
	\left|\frac{\delta M_{\mathrm{B}}}{M_{\mathrm{B}}}\right|\ll2\%\,,
	\qquad \delta M_{\mathrm{B}}<0\,.
\end{equation}
Here, $g$ is the effective quark-pair coupling used in the Hamiltonian. 

At NNLO, the effective three-quark interaction can receive contributions involving successive virtual propagation through the $4q\bar q$ and $5q2\bar q$ sectors before returning to the valence sector. The sign and magnitude of the combined correction depend on the corresponding transition matrix elements and intermediate-state energy denominators and therefore require an explicit calculation. 

Within the NLO estimate, the condition $M_{\mathrm B}<5\,m$ makes the five-parton denominator negative producing an attractive mass
shift. This is consistent with the ordering in Fig.~\ref{fig:MSU3}, where the valence masses lie above the DLCQ values~\cite{Hornbostel:1988fb}. More generally, for an equal-mass $N$-parton intermediate state, the condition $M_{\mathrm B}<N\,m$ makes the corresponding free invariant-mass denominator negative. Contributions with the same resolvent structure as the NLO term therefore produce an attractive shift relative to the LO ground-state mass. The increasing difference toward smaller $m/g_{\rm H}$ further supports the expectation that higher-Fock contributions become more important in the low-mass region. This
comparison neither determines the individual higher-Fock
contributions nor establishes the convergence pattern of the QP
expansion. Moreover, the LO wave function is normalized to unity within the valence
sector by construction. This normalization should not be interpreted
as the physical valence probability of the full state, which can be
determined only after the higher-Fock components are included.

\begin{figure}[t]
    \centering
    \includegraphics[width=0.48\textwidth]{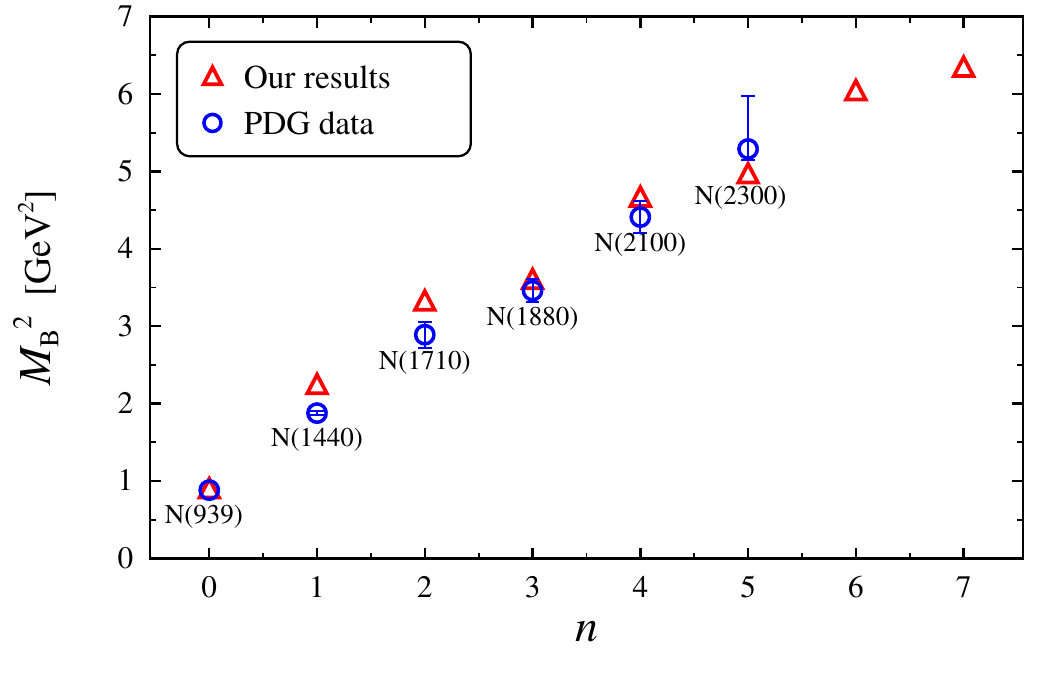}
    \caption{Regge trajectory of the squared nucleon masses. The model points correspond to the ground state and the first seven excited states, evaluated at $D=1000$. The accepted experimental values for the nucleon and its excited states are taken from the PDG~\cite{ParticleDataGroup:2024cfk}.}
    \label{fig:mass-spectra-N}
\end{figure}
\subsection{Regge trajectory}
The Regge trajectory for the radial excitations of the nucleon states ($J^P(I)=1/2^+(1/2)$) is shown in Fig.~\ref{fig:mass-spectra-N}. The calculated spectrum follows the overall trend of the experimental nucleon masses~\cite{ParticleDataGroup:2024cfk}, though it shows a slight deviation from a strictly linear trajectory. This feature appears to be an intrinsic property of the LO effective three-quark Hamiltonian and remains stable with increasing basis size for the lowest-lying states. Because the present model contains only longitudinal dynamics and
does not include transverse dynamics or realistic spin degrees of
freedom, the comparison with the experimental nucleon spectrum
should be understood as qualitative rather than quantitative.

We assess numerical stability by extending the basis calculation to
$D=1000$ for all states shown in Fig.~\ref{fig:mass-spectra-N}.
The ground-state mass converges rapidly and is stable to six decimal
places. Although the excited-state masses converge more slowly and
therefore require larger basis dimensions, at $D=1000$ they are
stable at the precision relevant to the displayed spectrum. Thus,
basis-truncation effects are negligible for both the ground-state
comparison and the excited-state spectrum shown here.

\subsection{Parton distribution functions}
Using the wave functions obtained in our model, we evaluate the unpolarized valence PDFs from the probability density in longitudinal momentum fraction,
\begin{equation}\label{eq:pdf}
	f^q(x)=n_q\int_0^{1-x} dy\,\big|\psi(x,y,1-x-y)\big|^2\,,
\end{equation}
where $q=u,d$ and $n_q$ denotes the number of valence quarks of
flavor $q$ in the nucleon. For the proton, $n_u=2$ and $n_d=1$.
The distributions satisfy the quark-number normalization conditions
\begin{equation}\label{eq:pdf_number_sum_rule}
    \int_0^1 dx\,f^q(x)=n_q.
\end{equation}
At the model scale, where the three valence quarks carry the entire
longitudinal momentum, they also satisfy the momentum sum rule
\begin{equation}\label{eq:pdf_momentum_sum_rule}
    \sum_{q=u,d}\int_0^1dx\,x f^q(x)=1.
\end{equation}

In the standard $(3+1)$-dimensional factorization framework, a PDF
depends on both the longitudinal momentum fraction $x$ and the
factorization scale $\mu$, after the transverse partonic momenta
have been integrated over~\cite{Collins:1981uw,Brodsky:1997de}.
We connect the two descriptions by treating the model distribution
$f^q(x)$ as the valence-PDF input at an effective matching scale
$\mu_0$. This identification is a phenomenological matching
assumption rather than an operator-level derivation.

We define the first moment of each valence distribution as $
    \langle x\rangle_q
    =\int_0^1dx\,x f^q(x)$.
The effective matching scale is fixed phenomenologically by requiring
the evolved total valence momentum fraction to satisfy $
    \langle x\rangle_u+\langle x\rangle_d
    \simeq 0.37$ at
    $\mu^2=10~\mathrm{GeV}^2,$
consistent with global PDF analyses~\cite{deTeramond:2018ecg}.
This procedure yields
$\mu_0^2=0.23\pm0.02~\mathrm{GeV}^2$. Because the total valence
momentum fraction is used to determine $\mu_0^2$, its agreement at
$\mu^2=10~\mathrm{GeV}^2$ is part of the calibration and does not
constitute an independent prediction.

The scale dependence applied after this matching comes from the (3+1)-dimensional DGLAP equations and not from QCD$_2$. Although the model distribution supplies the required longitudinal $x$ dependence, QCD$_2$ has a dimensionful coupling and no transverse degrees of freedom. It therefore does not generate the perturbative factorization structure or the splitting functions of (3+1)-dimensional QCD. We perform the evolution with HOPPET~\cite{Salam:2008qg} using the standard DGLAP equations~\cite{Dokshitzer:1977sg,Gribov:1972ri,Altarelli:1977zs}. This step is an external phenomenological procedure used to compare the model input with valence PDFs at higher scales. Starting from the low scale $\mu_0\simeq0.48~\mathrm{GeV}$ makes the result sensitive to the low-scale treatment of the evolution and to the matching prescription~\cite{Broniowski:2007si}. The quoted $10\%$ variation of $\mu_0^2$ measures only the sensitivity to the matching scale and does not represent a complete theory uncertainty.


The low-scale input inherits the parameter dependence of the
QCD$_2$ Hamiltonian. The mass parameter $m$ enters the quark
propagator and the kinetic term of the LO valence equation, while
the effective quark-pair coupling $g$, defined in
Eq.~\eqref{eq:coupling_dictionary}, controls the interaction
strength. We use $m=210~\mathrm{MeV}$ and
$g=330~\mathrm{MeV}$, as specified above. Because both parameters
have mass dimension one in QCD$_2$, the shape of the valence wave
function, and hence that of the low-scale PDF, depends on the ratio
$m/g$.

The present calculation does not impose a renormalization condition
relating the quark mass appearing in the QCD$_2$ Lagrangian to a
dressed mass. Establishing such a relation would require an explicit
self-energy or mass-gap analysis with a specified regularization
scheme~\cite{Ma:2021tHooft}. The value of $m$ used here should
therefore be regarded as a phenomenologically calibrated model
parameter within the LO valence truncation, rather than as a
$(3+1)$-dimensional $\overline{\mathrm{MS}}$ current-quark
mass. Its influence on the evolved PDFs enters through the
low-scale wave function and the resulting determination of
$\mu_0$.

\begin{figure}[t!]
	\centering
	\includegraphics[width=0.48\textwidth]{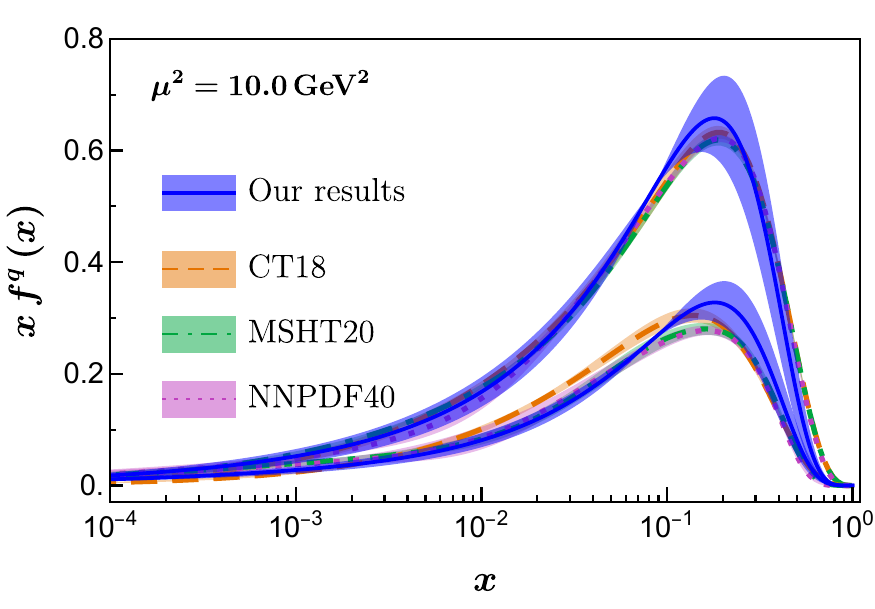}
	\caption{Unpolarized valence parton distribution functions of the proton, $x f^q(x)$, evolved to $\mu^2=10~\mathrm{GeV}^2$. The upper (lower) set of curves corresponds to the up- (down-) quark distribution. The blue bands indicate the uncertainty associated with the model, while the curves are compared with representative global analyses from the LHAPDF library, namely CT18, MSHT20, and NNPDF4.0~\cite{Buckley:2014ana}. }
	\label{fig:xpdf}
\end{figure}

The evolved valence PDFs at $\mu^2=10~\mathrm{GeV}^2$ are shown
in Fig.~\ref{fig:xpdf} and compared with the corresponding
$u_v$ and $d_v$ distributions from the CT18, MSHT20, and NNPDF4.0
global analyses, accessed through
LHAPDF~\cite{Buckley:2014ana}. Since the first moment is fixed by
the matching procedure, the remaining $x$ dependence provides a
qualitative test of the model distributions. The evolved
valence-PDF shapes are qualitatively consistent with the global
analyses.

\begin{figure*}[htp]
    \centering
\includegraphics[width=0.325\textwidth]{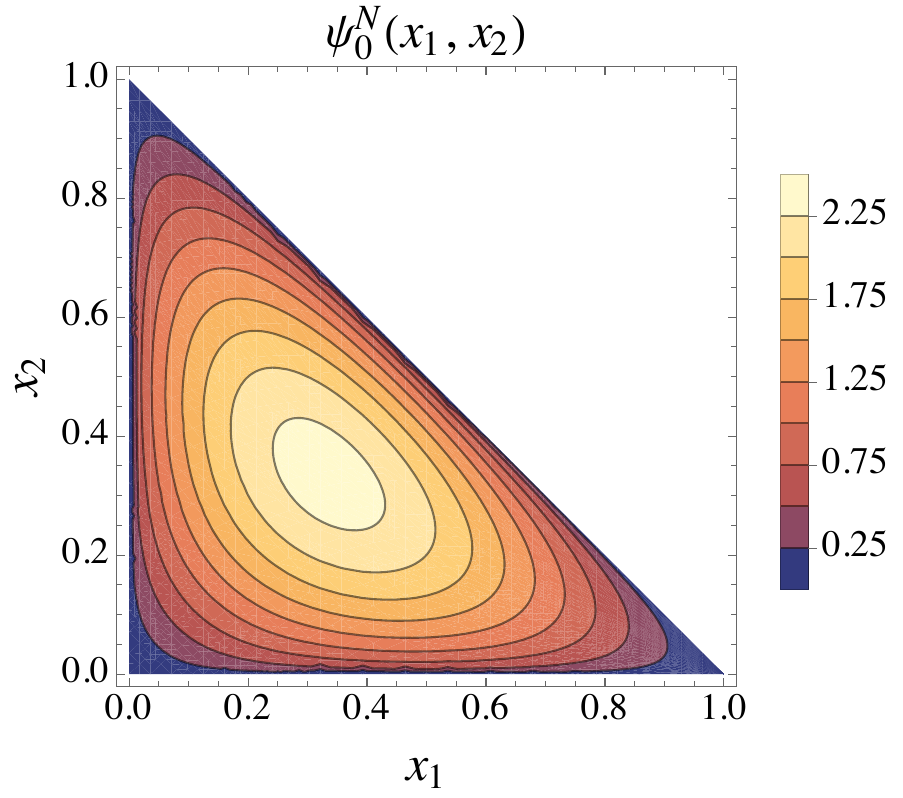} 
\includegraphics[width=0.325\textwidth]{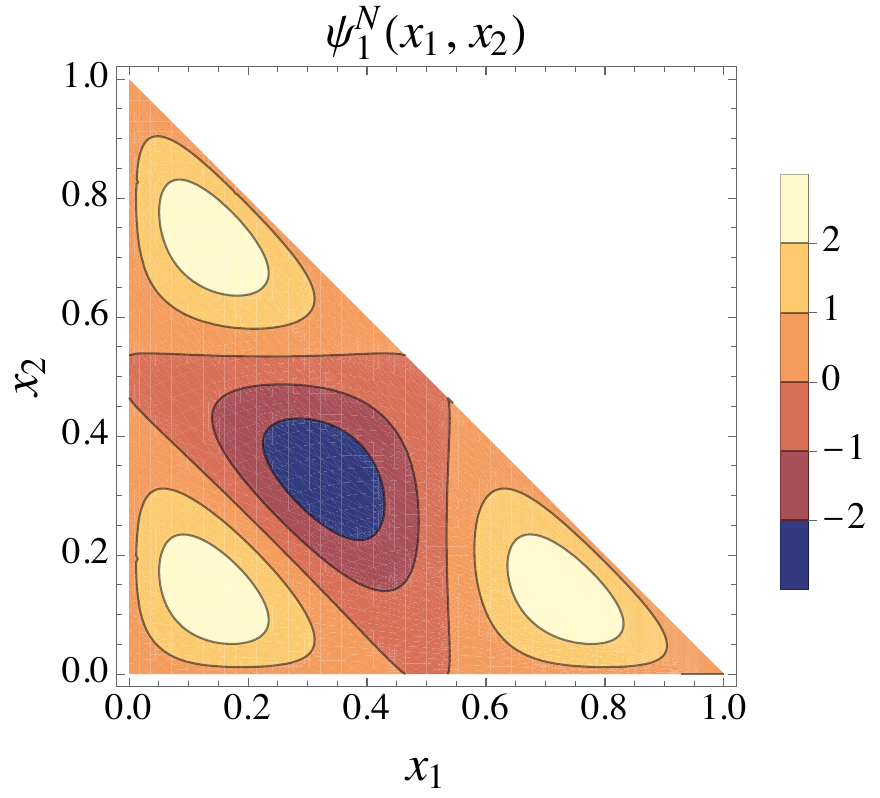} 
\includegraphics[width=0.325\textwidth]{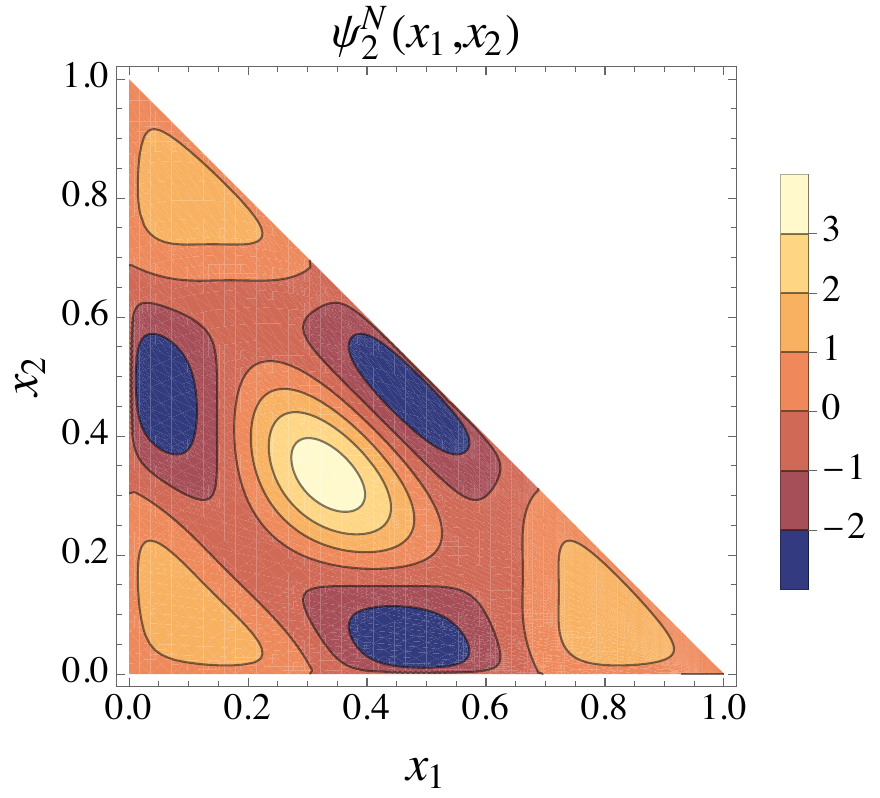} 
\includegraphics[width=0.325\textwidth]{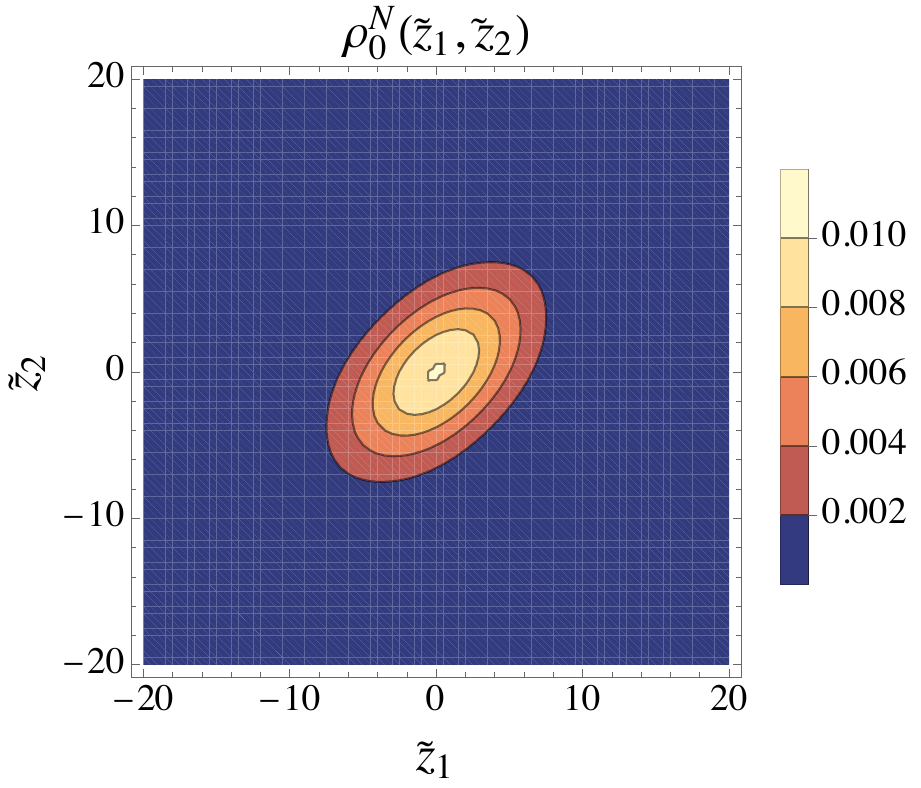}  
\includegraphics[width=0.325\textwidth]{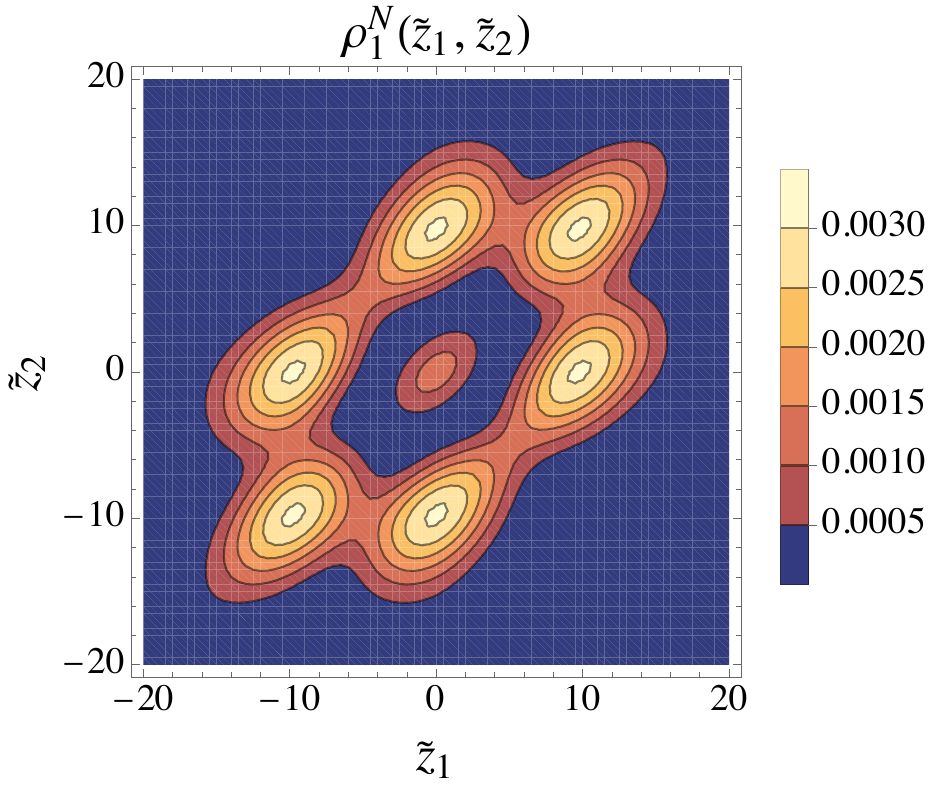}   
\includegraphics[width=0.325\textwidth]{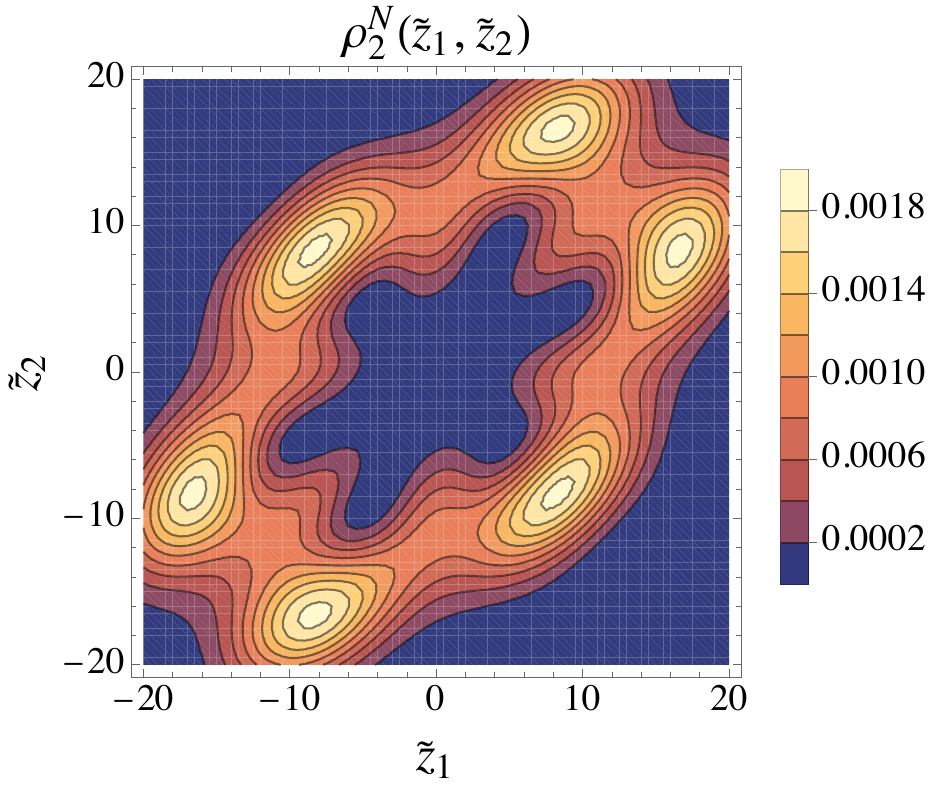}

    \caption{Top: Distribution amplitudes of the nucleon states: N(939) (left), N(1440) (middle) and N(1710) (right). Bottom: Co-ordinate space probability distribution: N(939) (left), N(1440) (middle) and N(1710) (right).}
    \label{fig:DA_1_2_3}
\end{figure*}

	\subsection{Double distribution amplitudes}
Using the resulting valence wave functions, the DDAs for the nucleon, $N(1440)$, and $N(1710)$ are shown in Fig.~\ref{fig:DA_1_2_3}~(top). The nucleon DDA exhibits the expected pattern, peaking at the symmetric point $x_i  = 1/3$ and vanishing at the domain boundaries with endpoint power-law behavior.

The Roper resonance, $N(1440)$, shows a different structure: the symmetric point becomes a global minimum, while maxima appear near the corners around $x_1 = x_2 \simeq 0.15$. The next resonance, $N(1710)$, displays a more complex pattern, retaining the central peak but with more pronounced corner maxima and an additional global minimum at $x_1 \simeq 0.5$, $x_2 \simeq 0.5$, while preserving overall symmetry.

In this model, the double parton distribution is given by the square of the DDA, so these features translate directly into a nontrivial quark momentum distribution. Notably, the DDA resembles that in Fig.~2 of Ref.~\cite{Ydrefors:2021mky}, where the valence wave function was obtained from the LO LF projection of the three-quark BSE with a contact interaction.

It should be emphasized that the present three-quark Hamiltonian is truncated at the valence level, so the valence eigenstates are mutually orthogonal. This property does not hold when higher Fock components are included, as the effective mass-squared operator then depends explicitly on the mass eigenvalue~\cite{Sales:1999ec}. In the exact solution of the BSE, Eq.~\eqref{eq:bse_1}, the direct relation between the DDAs and the double parton distributions is no longer exact, since the latter receives contributions from sectors beyond the valence~\cite{dePaula:2022pcb}. Finally, the overall patterns of the DDAs shown in Fig.~\ref{fig:DA_1_2_3} (top) for finite quark mass are qualitatively similar to those obtained by Webber for vanishing quark mass~\cite{Webber:1979na}.

	\subsection{Coordinate-space densities}
	
The coordinate-space probability distribution $\rho^N(\tilde z_1, \tilde z_2)$ for $N(939)$, $N(1440)$, and $N(1710)$ is shown in Fig.~\ref{fig:DA_1_2_3} (bottom). The dimensionless variables $\tilde z_1$ and $\tilde z_2$ correspond to the Ioffe-time coordinates introduced in Ref.~\cite{Miller:2019ysh}. The Fourier transform of the three-quark valence LF wave function is given by
\begin{equation}\label{eq:coordinater}
\hspace{-.3cm}		\tilde\psi(\tilde z_1,\tilde z_2)
		=
		\int_0^1 \hspace{-.1cm}\frac{dx_1}{\sqrt{2\pi}}
		\int_0^{1-x_1}\hspace{-.1cm} \frac{dx_2}{\sqrt{2\pi}}\,
		e^{i(x_1\tilde z_1+x_2\tilde z_2)}
		\,\psi(\{x_i\})\,,
	\end{equation}
	where $\{x_i\}\equiv\{x_1,x_2,x_3\}$ with $x_3=1-x_1-x_2$, and
	\begin{equation}
		\rho^N(\tilde z_1,\tilde z_2)=|\tilde\psi(\tilde z_1,\tilde z_2)|^2\,.
	\end{equation}
	
The ground-state density distribution, shown in the left panel of Fig.~\ref{fig:DA_1_2_3} (bottom), indicates that the quarks tend to minimize their relative separation in Ioffe time. This feature is also evident along  $\tilde z_1 = \tilde z_2$, which aligns with the major axis of the elliptical pattern. The density falls off away from this region due to the recoil of the third quark while keeping the center of mass at rest. No qualitative difference is observed compared to Fig.~3 of Ref.~\cite{Ydrefors:2021mky}. Together with the nucleon DDA, this suggests that the ground-state momentum- and coordinate-space distributions are relatively insensitive to interaction details, provided the basic scales, such as the nucleon and quark masses, are comparable.

The density distributions for $N(1440)$ and $N(1710)$, shown in the middle and right panels of Fig.~\ref{fig:DA_1_2_3} (bottom), display characteristic symmetric patterns in the $(\tilde z_1, \tilde z_2)$ plane, reflecting the structure of the corresponding DDAs. The spatial extent in this plane increases with the excitation of the nucleon state. Within the present valence-sector truncation, the valence wave functions remain mutually orthogonal.

\section{Summary and prospects}\label{sec:summ}
We performed the LF projection of the three-quark ladder BSE in Minkowski-space QCD$_2$ using the light-cone gauge. Employing the QP expansion, we derived the effective LF equation in the valence sector. After projection onto the color-singlet baryon, the quark-pair color
factor is absorbed into the effective coupling $g$, with
$g=\sqrt{2/3}\,g_{\rm H}$ for $N_c=3$. At LO, with the Fock-space truncated to the three-quark component, the projected equation reduces to the eigenvalue equation for an effective three-quark mass-squared operator, identical to the BDE~\cite{Bars:1976re,Bars:1976nk,Durgut:1976bc}. Following 't Hooft’s analysis of endpoint behavior in mesons~\cite{tHooft:1974pnl}, we derived a transcendental equation for the power-law exponent of the baryon valence wave function in terms of $m/g$. Beyond LO, the quasi-potential expansion generates irreducible three-body contributions from coupling to higher Fock components, consistent with studies of few-body BSEs in (3+1) dimensions~\cite{Marinho:2007zz,Marinho:2008zza,Karmanov:2009bhn,Ydrefors:2017nnc}.

The eigenvalue equation was solved numerically using a non-orthogonal, totally symmetric Jacobi-polynomial basis. For $N_c=3$, the calculated color-singlet ground-state masses agree closely with the independent valence calculation of Ref.~\cite{Zahed:2026qcd2} and lie above the central values of Hornbostel \textit{et al.}~\cite{Hornbostel:1988fb} at the nonzero tabulated ratios. This ordering is compatible with an attractive net contribution from the omitted higher Fock sectors, while a quantitative determination requires those sectors to be calculated explicitly. The same framework was applied to the nucleon Regge trajectory and various structure observables, including DDAs, coordinate-space probability densities in Ioffe-time coordinates, and valence parton distributions evolved to higher scales. The comparison with selected nucleon excitations is qualitative, and the evolved valence PDFs show encouraging agreement with global analyses at the level of their shapes.

The present results support using QCD$_2$ as a controlled confining test bed for Minkowski-space Bethe--Salpeter methods. Although substantial progress has been made in solving covariant BSEs with nonconfining kernels~\cite{Kusaka:1995za,Karmanov:2005nv,Frederico:2011ws,Carbonell:2014dwa,Ydrefors:2020duk,Eichmann:2021vnj}, extending these methods to explicitly confining theories remains challenging. In this context, the agreement between the present LO LF equation and light-cone QCD$_2$ results demonstrates that Minkowski-space approaches can capture the dominant features of confined baryons, providing a framework for future studies of baryon spectra and structure in QCD$_2$ and beyond.

\section*{Acknowledgments}
This work is supported by the National Natural Science Foundation of China (Grants No. 12305095, 12375143), the Gansu International Collaboration and Talents Recruitment Base of Particle Physics (2023–2027), and the Senior Scientist Program of Gansu Province (Grant No. 25RCKA008). JL is supported by the Special Research Assistant Funding Project and the Gansu Provincial Young Talents Program. SK is supported by the China Postdoctoral Science Foundation (Grant No. E339951SR0). XZ is supported by the Key Research Program of Frontier Sciences (Grant No. ZDBS-LY-7020), the International Partnership Program (Grant No. 016GJHZ2022103FN), and the Strategic Priority Research Program of the Chinese Academy of Sciences (Grant No. XDB34000000).
JPBCM and TF thank the Institute of Modern Physics (IMP) for their hospitality and the PIFI/CAS (Grants No. 2026PVA0083, 2026PVA0216) for financial support during their visit. They also acknowledge support from INCT-FNA (Grant No. 464898/2014-5) and FAPESP (Grants No. 2024/17816-8, 2023/09539-1). JPBCM additionally thanks CNPq (Grant No. 351403/2025-6). TF also acknowledges support from CNPq (Grant No. 306834/2022-7) and FAPESP (Grant No. 2023/13749-1 and 2025/05312-8).

\paragraph{Note added}
After the first version of this work appeared, a complementary study of baryons in QCD$_2$ with
$N_c=3$, based on a light-front Faddeev formulation, was reported in Ref.~\cite{Zahed:2026qcd2}. The authors developed a coupled-channel
representation of the three-quark problem and solved the resulting
valence-sector equation variationally, with applications to the
chiral spectrum, isospin, and strangeness. In the region of overlap,
their ground-state baryon masses agree closely with our results, as
shown in Fig.~\ref{fig:MSU3} and
Table~\ref{tab:ground_state_comparison}.

\appendix

\section{LF projection of the baryon BSE}\label{sec:LFBSE}

To introduce the LF projection in a transparent way, we first consider the corresponding three-boson Bethe--Salpeter amplitude,
\begin{equation}\label{Phi0}
	\Phi_M^b(y_1,y_2,y_3;K)=
	\langle 0 | T\big(\varphi(y_1)\varphi(y_2)\varphi(y_3)\big) |K\rangle \,,
\end{equation}
where $y_i$ denotes the spacetime coordinate of particle $i$, $\varphi(y)$ is the bosonic field operator, and $K$ is the total momentum. The superscript $b$ indicates the bosonic system.

In (1+1) dimensions, the valence LF wave function is obtained by projecting the BS amplitude onto the LF,
\begin{equation}\label{bs7b}
	\psi_3^b(x_1,x_2,x_3)=
	(K^+)^2 (x_1x_2x_3)^{1/2}\,\chi_3^b(x_1,x_2,x_3)\,,
\end{equation}
where
\begin{equation}\label{bs7c}
	\chi_3^b(x_1,x_2,x_3)=
	\int dk_1^-\,dk_2^-\,\Phi_M^b(k_1,k_2,k_3;K)\,.
\end{equation}
The integration over the minus components removes the relative LF time among the three particles. Here $\Phi_M^b$ denotes the momentum-space representation of the Minkowski-space Bethe--Salpeter amplitude, and $\chi_3^b$ is introduced as an auxiliary LF amplitude.

For any operator $A$ defined in Minkowski space, we represent the LF projection by
\begin{equation}\label{bardef3}
\begin{aligned}
	|A := \int dk_1^- dk_2^-\, \langle k_1^-,k_2^-|A\,, \\
	A| := \int dk_1^- dk_2^-\, A|k_1^-,k_2^-\rangle \,.
    \end{aligned}
\end{equation}
After factoring out the total momentum, the matrix elements of $A$ depend only on two independent momenta. Using this notation, the projected three-body amplitude and the corresponding vertex function satisfy
\begin{equation}
	|\chi_3^b\rangle = |G_0^b|\Gamma_M^b\rangle \,,
	\qquad
	|\Gamma_M^b\rangle = V^b G_0^b |\Gamma_M^b\rangle \,,
\end{equation}
respectively, where $V^b$ is the interaction kernel, and the second relation is the BSE for the vertex function.

The disconnected three-boson Green's function is
\begin{multline}\label{eq:g03boson}
	\hspace{-.1cm}	\langle k_1^-,k_2^-|G_0^b|k_1^{\prime -},k_2^{\prime -}\rangle
		=
		\frac{-i}{(2\pi)^2}
		\frac{\delta(k_1^- - k_1^{\prime -})\,
			\delta(k_2^- - k_2^{\prime -})}
		{\hat{k}_1^+\,\hat{k}_2^+\,(K^+-\hat{k}_1^+-\hat{k}_2^+)}
		\\ \hspace{-.15cm}\times
		\frac{1}
		{(k_1^- - \hat{k}_{1\mathrm{on}}^-)
			(k_2^- - \hat{k}_{2\mathrm{on}}^-)
			(K^- - k_1^- - k_2^- - \hat{k}_{3\mathrm{on}}^-)}\,.
	\end{multline}
Here the on-shell minus component is
$	\hat{k}_{i\mathrm{on}}^-=m_i^2/\hat{k}_i^+$,
and, for notational simplicity, we understand
$	\hat{k}_{i\mathrm{on}}^- \to \hat{k}_{i\mathrm{on}}^- - {i\varepsilon}/{\hat{k}_i^+}$
to implement the standard causal prescription. Momentum conservation implies
$	\hat{k}_3^+ = K^+ - \hat{k}_1^+ - \hat{k}_2^+$,
with the corresponding relation holding for the remaining momentum components.

\subsection{QP expansion for baryons}

We now return to the three-quark system. The disconnected three-quark Green's function can be expressed in terms of the  bosonic Green's function~\cite{Ydrefors:2021mky,Ydrefors:2022bhq} as
\begin{multline}\label{eq:g03b}
	\langle k_1^-,k_2^- |G_0|k_1^{\prime -},k_2^{\prime -}\rangle
	=
	\langle k_1^-,k_2^- |G_0^b|k_1^{\prime -},k_2^{\prime -}\rangle
	\\ \times
	(\psla k_{1}+m)\otimes(\psla k_{2}+m)\otimes(\psla k_{3}+m)\,.
\end{multline}

To perform the LF projection of the BSE, the instantaneous contribution in each quark propagator must first be separated, leaving only the propagating part~\cite{Sales:2001gk}. This procedure defines
\begin{multline}\label{eq:g03bb}
	\langle k_1^-,k_2^- |\overline G_0|k_1^{\prime -},k_2^{\prime -}\rangle
	=
	\langle k_1^-,k_2^- |G_0^b|k_1^{\prime -},k_2^{\prime -}\rangle
	\\ \times
	(\psla k_{1\mathrm{on}}+m)\otimes(\psla k_{2\mathrm{on}}+m)\otimes(\psla k_{3\mathrm{on}}+m)\,.
\end{multline}

The LF projection of Eq.~\eqref{eq:g03bb},
$	g_0 = |\overline G_0|$,
defines the free LF resolvent. Explicitly,
\begin{multline}\label{propfl3}
	g_0(k_1^+,k_2^+,k_3^+)
	=
	(2m)^3\,g_0^b(k_1^+,k_2^+,k_3^+)
	\\ \times
	\Lambda^+(k_1^+)\otimes\Lambda^+(k_2^+)\otimes\Lambda^+(k_3^+)\,,
\end{multline}
where
$	\Lambda^+(k^+) = {(\psla k_{\mathrm{on}}+m)}/{2m}$
is the Dirac projector, and the bosonic-like resolvent is
\begin{multline}\label{propfl31}
	g_0^b(k_1^+,k_2^+,k_3^+)
	=
	\frac{i}{k_1^+k_2^+(K^+-k_1^+-k_2^+)}
	\\ \times
	\frac{\theta(K^+-k_1^+-k_2^+)\,\theta(k_1^+)\,\theta(k_2^+)}
	{K^- - k_{1\mathrm{on}}^- - k_{2\mathrm{on}}^- - (K-k_1-k_2)_{\mathrm{on}}^-}\,.
\end{multline}

To perform the LF reduction, we employ the QP method~\cite{Sales:1999ec,Sales:2001gk}, in which an auxiliary Green's function is introduced:
$	\widetilde{G}_0 = \overline G_0|g_0^{-1}|\overline G_0$.
The four-dimensional BSE for the vertex function can then be expressed as
\begin{equation}
	|\Gamma_M\rangle
	=
	W \widetilde G_0 |\Gamma_M\rangle
	=
	W \overline G_0|g_0^{-1}|\overline G_0|\Gamma_M\rangle\,.
\end{equation}
The QP satisfies $W = V + V\Delta_0 W$
with $\Delta_0 = G_0 - \widetilde G_0$.

The LF valence amplitude therefore follows
\begin{equation}\label{eq:LFprojBS}
	|\chi_3\rangle
	=
	|\overline G_0|\Gamma_M\rangle
	=
	g_0\,w\,|\overline G_0|\Gamma_M\rangle
	=
	g_0\,w\,|\chi_3\rangle\,,
\end{equation}
where the effective LF interaction acting in the valence sector is $w = g_0^{-1}|\overline G_0 W \overline G_0|g_0^{-1}$.

The interaction kernel is constructed from the pairwise instantaneous gluon exchanges between quarks and can be expressed as
\begin{equation}\label{Eq:V}
	V=\sum_{i=1}^3 V_i
	=
	\sum_{i=1}^3 S_i^{-1}\otimes V_{(2)i}\,,
\end{equation}
where $V_{(2)i}$ denotes the two-body interaction between the pair of quarks different from $i$.

The Faddeev decomposition of the QP and the corresponding LF interaction is given by
\begin{equation}\label{Eq:Faddeev_W}
	W=\sum_i W_i\,,
	\qquad
	w=\sum_i w_i\,,
\end{equation}
with $w_i = g_0^{-1}|\overline G_0 W_i \overline G_0|g_0^{-1}$.
The Faddeev components admit the expansion
\begin{equation}\label{Eq:W_exp}
	\begin{aligned}
		W_i
		&= V_i + V_i\Delta_0(V_i+V_j+V_k) \\
		&\quad + V_i\Delta_0(V_i+V_j+V_k)\Delta_0(V_i+V_j+V_k)+\cdots \, .
	\end{aligned}
\end{equation}
At LO, the effective interaction for the valence component of the baryon LF wave function reduces to
\begin{equation}\label{Eq:w_L1}
	w_i^{\mathrm{LO}} = g_0^{-1}|\overline G_0 V_i \overline G_0|g_0^{-1}\,.
\end{equation}

Beyond LO, the QP expansion generates additional contributions to the effective interaction through the coupling of the valence component to higher Fock sectors. In particular, an NLO irreducible three-body contribution to the effective valence-space Hamiltonian is (see Fig.~\ref{fig:kernel_LO}, right):
\begin{equation}
    \delta w^{\mathrm{NLO}}_i= g_0^{-1}|\overline G_0 V_i\Delta_0(V_j+V_k) \overline G_0|g_0^{-1}\,,
    \label{Eq:w_NLO1}
\end{equation}
 where $|\overline G_0V_i\Delta_0V_i\overline G_0|$ vanishes after the LF projection, namely the integration over the minus momentum of the quarks, because the gluon exchange is instantaneous in the light-cone gauge, and the quark-gluon coupling has only $\gamma^+$, which in addition eliminates the instantaneous contributions from the quark propagators.

Here, we restrict ourselves to the LO interaction, Eq.~\eqref{Eq:w_L1}, derived below.

	\subsection{Derivation of $w_i^{\mathrm{LO}}$}
	
	We now evaluate the LF projected operator $|\overline G_0 V_i \overline G_0|$ entering the LO effective interaction in Eq.~\eqref{Eq:w_L1}. Without loss of generality, we choose $i=1$. For compactness, we define
	\begin{equation}
		\mathcal{N}_3(k_1,k_2,k_3)
		\equiv \prod_{i=1,3}
		(\psla k_{i\mathrm{on}}+m)\otimes\,,
	\end{equation}
and then we write:
 	\begin{align}
		&|\overline G_0 V_1 \overline G_0|
		=
		-\frac{1}{(2\pi)^4}
		\frac{1}{\hat{k}_1^+\hat{k}_2^+(K^+-\hat{k}_1^+-\hat{k}_2^+)}
		\int dk_1^- dk_2^- dk_2^{\prime -}\nonumber\\
		 &\times\frac{\mathcal{N}_3(k_1,k_2,k_3)}
		{(k_1^- - \hat{k}_{1\mathrm{on}}^-)(k_2^- - \hat{k}_{2\mathrm{on}}^-)(K^- - k_1^- - k_2^- - \hat{k}_{3\mathrm{on}}^-)}\nonumber\\
		 &\times
		\frac{\gamma_1^+\otimes\gamma_2^+\otimes\gamma_3^+}{4(k_2^+-k_2^{\prime +})^2}
		\frac{1}{\hat{k}_1^+\hat{k}_2^{\prime +}(K^+-\hat{k}_1^+-\hat{k}_2^{\prime +})}\nonumber\\
		 &\times\frac{\mathcal{N}_3(k_1,k_2',k_3')}
		{(k_2^{\prime -} - \hat{k}_{2\mathrm{on}}^{\prime -})(K^- - k_1^- - k_2^{\prime -} - \hat{k}_{3\mathrm{on}}^{\prime -})}\, ,
	\end{align}   
where  $1/2$ comes from the Jacobian $(d^2k=\frac12\,dk^+dk^-)$, and the identities: $S_1^{-1}(k_1)=\psla k_{1\mathrm{on}}-m+{\gamma_1^+}(k_1^- - \hat{k}_{1\mathrm{on}}^-)/{2}$
	and 
    $(\psla k_{1\mathrm{on}}-m)({\psla k}_{1\mathrm{on}}+m)=0$ were taken into account.

Using Cauchy integration and the causal prescription $\hat{k}_{i\mathrm{on}}^- \to \hat{k}_{i\mathrm{on}}^- - {i\varepsilon}/{\hat{k}_i^+}$,
the pairwise LO effective interaction acting on the baryon LF valence state becomes
\begin{equation}
	\label{Eq:w_LP4}
	w_i^{\mathrm{LO}}
	=
	\frac{i}{8\pi}
	\frac{\gamma_i^+\otimes\gamma_j^+\otimes\gamma_k^+}{(k_j^+ - k_j^{\prime +})^2}\,.
\end{equation}
This expression corresponds to the instantaneous gluon exchange between quarks $j$ and $k$.

\biboptions{sort&compress}
\bibliographystyle{elsarticle-num}
\bibliography{references_cleaned}

@article{tHooft:1974pnl,
    author = "'t Hooft, Gerard",
    title = "{A Two-Dimensional Model for Mesons}",
    reportNumber = "CERN-TH-1820",
    doi = "10.1016/0550-3213(74)90088-1",
    journal = "Nucl. Phys. B",
    volume = "75",
    pages = "461--470",
    year = "1974"
}

@article{Hornbostel:1988fb,
    author = "Hornbostel, Kent and Brodsky, Stanley J. and Pauli, Hans Christian",
    title = "{Light Cone Quantized QCD in (1+1)-Dimensions}",
    reportNumber = "SLAC-PUB-4678",
    doi = "10.1103/PhysRevD.41.3814",
    journal = "Phys. Rev. D",
    volume = "41",
    pages = "3814",
    year = "1990"
}

@article{Zahed:2026qcd2,
    author = "Zahed, Ismail and Zhang, Haocheng",
    title = "{Faddeev description of baryons in two-dimensional QCD with $N_c=3$. I. Chiral spectrum, isospin, and strangeness}",
    eprint = "2607.23435",
    archivePrefix = "arXiv",
    primaryClass = "hep-ph",
    month = "7",
    year = "2026"
}

@article{Vary:2009gt,
    author = "Vary, J. P. and Honkanen, H. and Li, Jun and Maris, P. and Brodsky, S. J. and Harindranath, A. and de Teramond, G. F. and Sternberg, P. and Ng, E. G. and Yang, C.",
    title = "{Hamiltonian light-front field theory in a basis function approach}",
    eprint = "0905.1411",
    archivePrefix = "arXiv",
    primaryClass = "nucl-th",
    reportNumber = "SLAC-PUB-13582",
    doi = "10.1103/PhysRevC.81.035205",
    journal = "Phys. Rev. C",
    volume = "81",
    pages = "035205",
    year = "2010"
}

@article{Vary:2025yqo,
    author = "Vary, James P. and Mondal, Chandan and Xu, Siqi and Zhao, Xingbo and Li, Yang",
    title = "{Nucleon Structure from Basis Light-Front Quantization : Status and Prospects}",
    eprint = "2512.08283",
    archivePrefix = "arXiv",
    primaryClass = "hep-ph",
    doi = "10.1140/epjs/s11734-025-02084-y",
    month = "12",
    year = "2025"
}

@article{Nakarev,
      author         = "Nakanishi, Noboru",
      title          = "{A General survey of the theory of the Bethe-Salpeter
                        equation}",
      journal        = "Prog.~Theor.~Phys.~Suppl.",
      volume         = "43",
      year           = "1969",
      pages          = "1-81",
      doi            = "10.1143/PTPS.43.1",
      SLACcitation   = "%%CITATION = PTPSA,43,1;%%"
}

@article{Kusaka:1995za,
    author = "Kusaka, Kensuke and Williams, Anthony G.",
    title = "{Solving the Bethe-Salpeter equation for scalar theories in Minkowski space}",
    eprint = "hep-ph/9501262",
    archivePrefix = "arXiv",
    reportNumber = "ADP-94-24-T-164, ADP-94-24-T164",
    doi = "10.1103/PhysRevD.51.7026",
    journal = "Phys.~Rev.~D",
    volume = "51",
    pages = "7026--7039",
    year = "1995"
}

@article{Karmanov:2005nv,
      author         = "Karmanov, V. A. and Carbonell, J.",
      title          = "{Solving Bethe-Salpeter equation in Minkowski space}",
      journal        = "Eur. Phys. J. A",
      volume         = "27",
      year           = "2006",
      pages          = "1-9",
      doi            = "10.1140/epja/i2005-10193-0",
      eprint         = "hep-th/0505261",
      archivePrefix  = "arXiv",
      primaryClass   = "hep-th",
      SLACcitation   = "%%CITATION = HEP-TH/0505261;%%"
}

@article{Frederico:2011ws,
    author = "Frederico, Tobias and Salme, Giovanni and Viviani, Michele",
    title = "{Two-body scattering states in Minkowski space and the Nakanishi integral representation onto the null plane}",
    eprint = "1112.5568",
    archivePrefix = "arXiv",
    primaryClass = "hep-ph",
    doi = "10.1103/PhysRevD.85.036009",
    journal = "Phys. Rev. D",
    volume = "85",
    pages = "036009",
    year = "2012"
}

@article{Ydrefors:2020duk,
    author = "Ydrefors, E. and Alvarenga Nogueira, J. H. and Karmanov, V. A. and Frederico, T.",
    title = "{Three-boson bound states in Minkowski space with contact interactions}",
    eprint = "2005.07943",
    archivePrefix = "arXiv",
    primaryClass = "hep-ph",
    doi = "10.1103/PhysRevD.101.096018",
    journal = "Phys. Rev. D",
    volume = "101",
    number = "9",
    pages = "096018",
    year = "2020"
}

@article{Carbonell:2014dwa,
    author = "Carbonell, J. and Karmanov, V. A.",
    title = "{Solving Bethe-Salpeter scattering state equation in Minkowski space}",
    eprint = "1408.3761",
    archivePrefix = "arXiv",
    primaryClass = "hep-ph",
    doi = "10.1103/PhysRevD.90.056002",
    journal = "Phys. Rev. D",
    volume = "90",
    number = "5",
    pages = "056002",
    year = "2014"
}

@article{Eichmann:2021vnj,
    author = "Eichmann, Gernot and Ferreira, Eduardo and Stadler, Alfred",
    title = "{Going to the light front with contour deformations}",
    eprint = "2112.04858",
    archivePrefix = "arXiv",
    primaryClass = "hep-ph",
    doi = "10.1103/PhysRevD.105.034009",
    journal = "Phys. Rev. D",
    volume = "105",
    number = "3",
    pages = "034009",
    year = "2022"
}

@article{dePaula:2026gtb,
    author = "de Paula, Wayne and Frederico, Tobias",
    title = "{Minkowski Space Dynamics and Light-Front Projection}",
    eprint = "2601.11760",
    archivePrefix = "arXiv",
    primaryClass = "hep-ph",
    month = "1",
    year = "2026"
}

@article{Li:2021jqb,
	author = "Li, Yang and Vary, James P.",
	title = "{Light-front holography with chiral symmetry breaking}",
	eprint = "2103.09993",
	archivePrefix = "arXiv",
	primaryClass = "hep-ph",
	doi = "10.1016/j.physletb.2021.136860",
	journal = "Phys. Lett. B",
	volume = "825",
	pages = "136860",
	year = "2022"
}

@article{Li2022,
    author = "Li, Yang and Maris, Pieter and Vary, James P.",
    title = "{Chiral sum rule on the light front and the 3D image of the pion}",
    eprint = "2203.14447",
    archivePrefix = "arXiv",
    primaryClass = "hep-th",
    doi = "10.1016/j.physletb.2022.137598",
    journal = "Phys. Lett. B",
    volume = "836",
    pages = "137598",
    year = "2023"
}

@article{deTeramond:2021yyi,
    author = "de T\'eramond, Guy F. and Brodsky, Stanley J.",
    title = "{Longitudinal dynamics and chiral symmetry breaking in holographic light-front QCD}",
    eprint = "2103.10950",
    archivePrefix = "arXiv",
    primaryClass = "hep-ph",
    reportNumber = "SLAC-PUB-17593",
    doi = "10.1103/PhysRevD.104.116009",
    journal = "Phys. Rev. D",
    volume = "104",
    number = "11",
    pages = "116009",
    year = "2021"
}

@article{Lyubovitskij:2022rod,
    author = "Lyubovitskij, Valery E. and Schmidt, Ivan",
    title = "{Meson masses and decay constants in holographic QCD consistent with ChPT and HQET}",
    eprint = "2203.00604",
    archivePrefix = "arXiv",
    primaryClass = "hep-ph",
    doi = "10.1103/PhysRevD.105.074009",
    journal = "Phys. Rev. D",
    volume = "105",
    number = "7",
    pages = "074009",
    year = "2022"
}

@article{Weller:2021wog,
    author = "Weller, Colin M. and Miller, Gerald A.",
    title = "{Confinement in two-dimensional QCD and the infinitely long pion}",
    eprint = "2111.03194",
    archivePrefix = "arXiv",
    primaryClass = "hep-ph",
    doi = "10.1103/PhysRevD.105.036009",
    journal = "Phys. Rev. D",
    volume = "105",
    number = "3",
    pages = "036009",
    year = "2022"
}

@article{Ahmady:2021lsh,
    author = "Ahmady, Mohammad and Dahiya, Harleen and Kaur, Satvir and Mondal, Chandan and Sandapen, Ruben and Sharma, Neetika",
    title = "{Extending light-front holographic QCD using the 't Hooft Equation}",
    eprint = "2105.01018",
    archivePrefix = "arXiv",
    primaryClass = "hep-ph",
    doi = "10.1016/j.physletb.2021.136754",
    journal = "Phys. Lett. B",
    volume = "823",
    pages = "136754",
    year = "2021"
}

@article{Rinaldi:2022dyh,
    author = "Rinaldi, Matteo and Ceccopieri, Federico Alberto and Vento, Vicente",
    title = "{The pion in the graviton soft-wall model: phenomenological applications}",
    eprint = "2204.09974",
    archivePrefix = "arXiv",
    primaryClass = "hep-ph",
    doi = "10.1140/epjc/s10052-022-10538-z",
    journal = "Eur. Phys. J. C",
    volume = "82",
    number = "7",
    pages = "626",
    year = "2022"
}

@article{Gurjar:2025kcp,
    author = "Gurjar, Bheemsehan and Mondal, Chandan and Kaur, Satvir",
    title = {{{\ensuremath{\phi}}-meson spectroscopy and diffractive production using two Schr{\"o}dinger-like equations on the light front}},
    eprint = "2501.11436",
    archivePrefix = "arXiv",
    primaryClass = "hep-ph",
    doi = "10.1103/PhysRevD.111.094002",
    journal = "Phys. Rev. D",
    volume = "111",
    number = "9",
    pages = "094002",
    year = "2025"
}

@article{Kaur:2025css,
    author = "Kaur, Satvir and Mondal, Chandan and Zhao, Xingbo and Ji, Chueng-Ryong",
    title = "{Structure of the lightest nucleus in the visible Universe}",
    eprint = "2507.09886",
    archivePrefix = "arXiv",
    primaryClass = "hep-ph",
    doi = "10.1103/xkdk-ymn6",
    journal = "Phys. Rev. D",
    volume = "113",
    number = "5",
    pages = "054008",
    year = "2026"
}

@article{Kaur:2025gyr,
    author = "Kaur, Satvir and Mondal, Chandan",
    title = "{Gluon distributions in the pion}",
    eprint = "2507.01506",
    archivePrefix = "arXiv",
    primaryClass = "hep-ph",
    doi = "10.1103/j5x3-ljwd",
    journal = "Phys. Rev. D",
    volume = "112",
    number = "11",
    pages = "114015",
    year = "2025"
}

@article{Gurjar:2024wpq,
    author = "Gurjar, Bheemsehan and Mondal, Chandan and Kaur, Satvir",
    title = {{\ensuremath{\rho}-meson spectroscopy and diffractive production using the holographic light-front Schr\"odinger equation and the \textquoteright{}t Hooft equation}},
    eprint = "2401.13514",
    archivePrefix = "arXiv",
    primaryClass = "hep-ph",
    doi = "10.1103/PhysRevD.109.094017",
    journal = "Phys. Rev. D",
    volume = "109",
    number = "9",
    pages = "094017",
    year = "2024"
}

@article{Ahmady:2022dfv,
    author = "Ahmady, Mohammad and Kaur, Satvir and Mondal, Chandan and Sandapen, Ruben",
    title = {{Pion spectroscopy and dynamics using the holographic light-front Schr\"odinger equation and the 't Hooft equation}},
    eprint = "2208.08405",
    archivePrefix = "arXiv",
    primaryClass = "hep-ph",
    doi = "10.1016/j.physletb.2022.137628",
    journal = "Phys. Lett. B",
    volume = "836",
    pages = "137628",
    year = "2023"
}

@article{Ahmady:2021yzh,
    author = "Ahmady, Mohammad and Kaur, Satvir and MacKay, Sugee Lee and Mondal, Chandan and Sandapen, Ruben",
    title = {{Hadron spectroscopy using the light-front holographic Schr\"odinger equation and the \textquoteright{}t Hooft equation}},
    eprint = "2108.03482",
    archivePrefix = "arXiv",
    primaryClass = "hep-ph",
    doi = "10.1103/PhysRevD.104.074013",
    journal = "Phys. Rev. D",
    volume = "104",
    number = "7",
    pages = "074013",
    year = "2021"
}

@article{Bars:1976re,
    author = "Bars, Itzhak",
    title = "{Exact Equivalence of Chromodynamics to a String Theory}",
    reportNumber = "COO-3075-140",
    doi = "10.1103/PhysRevLett.36.1521",
    journal = "Phys. Rev. Lett.",
    volume = "36",
    pages = "1521",
    year = "1976"
}

@article{Bars:1976nk,
    author = "Bars, I.",
    title = "{A Quantum String Theory of Hadrons and Its Relation to Quantum Chromodynamics in Two-Dimensions}",
    reportNumber = "COO-3075-142",
    doi = "10.1016/0550-3213(76)90327-8",
    journal = "Nucl. Phys. B",
    volume = "111",
    pages = "413--440",
    year = "1976"
}

@article{Durgut:1976bc,
    author = "Durgut, Metin",
    title = "{Baryon Bound State in Two-Dimensional SU(N) Gauge Theory}",
    reportNumber = "ITP-SB-76-18",
    doi = "10.1016/0550-3213(76)90324-2",
    journal = "Nucl. Phys. B",
    volume = "116",
    pages = "233--252",
    year = "1976"
}

@article{Webber:1979na,
    author = "Webber, B. R.",
    title = "{Solution of a Two-dimensional {QCD} Model for Baryons}",
    reportNumber = "HEP 79/1",
    doi = "10.1016/0550-3213(79)90609-6",
    journal = "Nucl. Phys. B",
    volume = "153",
    pages = "455--466",
    year = "1979"
}

@article{Sales:1999ec,
      author         = "Sales, J. H. O. and Frederico, T. and Carlson, B. V. and
                        Sauer, P. U.",
      title          = "{Light front Bethe-Salpeter equation}",
      journal        = "Phys.~Rev.~C",
      volume         = "61",
      year           = "2000",
      pages          = "044003",
      doi            = "10.1103/PhysRevC.61.044003",
      eprint         = "nucl-th/9909029",
      archivePrefix  = "arXiv",
      primaryClass   = "nucl-th",
      SLACcitation   = "%%CITATION = NUCL-TH/9909029;%%"
}

@article{Sales:2001gk,
      author         = "Sales, J. H. O. and Frederico, T. and Carlson, B. V. and
                        Sauer, P. U.",
      title          = "{Renormalization of the ladder light front Bethe-Salpeter
                        equation in the Yukawa model}",
      journal        = "Phys.~Rev.~C",
      volume         = "63",
      year           = "2001",
      pages          = "064003",
      doi            = "10.1103/PhysRevC.63.064003",
      SLACcitation   = "%%CITATION = PHRVA,C63,064003;%%"
}

@article{Marinho:2007zz,
    author = "Marinho, J. A. O. and Frederico, T.",
    editor = "Mathiot, Jean-Francois and Bakker, Bernard and Diehl, Markus",
    title = "{Next-to-leading order light-front three-body dynamics}",
    doi = "10.22323/1.061.0036",
    journal = "PoS",
    volume = "LC2008",
    pages = "036",
    year = "2008"
}

@article{Marinho:2008zza,
    author = "Marinho, J. A. O. and Frederico, T.",
    editor = "Barlow, Roger",
    title = "{Three-boson systems in light-front dynamics}",
    doi = "10.1088/1742-6596/110/12/122009",
    journal = "J. Phys. Conf. Ser.",
    volume = "110",
    pages = "122009",
    year = "2008"
}

@article{Guimaraes:2014kor,
    author = "Guimar\~aes, K. S. F. F. and Louren\c{c}o, O. and de Paula, W. and Frederico, T. and dos Reis, A. C.",
    title = "{Final state interaction in $D^{+} \to K^{-} \pi^{+} \pi^{+}$ with $K\pi$ I = 1/2 and 3/2 channels}",
    eprint = "1404.3797",
    archivePrefix = "arXiv",
    primaryClass = "hep-ph",
    doi = "10.1007/JHEP08(2014)135",
    journal = "{JHEP}",
    volume = "{08}",
    pages = "{135}",
    year = "2014"
}

@article{Ydrefors:2021mky,
    author = "Ydrefors, Emanuel and Frederico, Tobias",
    title = "{Proton image and momentum distributions from light-front dynamics}",
    eprint = "2108.02146",
    archivePrefix = "arXiv",
    primaryClass = "hep-ph",
    doi = "10.1103/PhysRevD.104.114012",
    journal = "Phys. Rev. D",
    volume = "104",
    number = "11",
    pages = "114012",
    year = "2021"
}

@article{Brodsky:1997de,
    author = "Brodsky, Stanley J. and Pauli, Hans-Christian and Pinsky, Stephen S.",
    title = "{Quantum chromodynamics and other field theories on the light cone}",
    eprint = "hep-ph/9705477",
    archivePrefix = "arXiv",
    reportNumber = "SLAC-PUB-7484, MPIH-V1-1997",
    doi = "10.1016/S0370-1573(97)00089-6",
    journal = "Phys. Rept.",
    volume = "301",
    pages = "299--486",
    year = "1998"
}

@article{Mo:1992sv,
    author = "Mo, Yi-zhang and Perry, Robert J.",
    title = "{Basis function calculations for the massive Schwinger model in the light front Tamm-Dancoff approximation}",
    reportNumber = "OSU-NT-92-128",
    doi = "10.1006/jcph.1993.1171",
    journal = "J. Comput. Phys.",
    volume = "108",
    pages = "159--174",
    year = "1993"
}

@article{ParticleDataGroup:2024cfk,
    author = "Navas, S. and others",
    collaboration = "Particle Data Group",
    title = "{Review of particle physics}",
    doi = "10.1103/PhysRevD.110.030001",
    journal = "Phys. Rev. D",
    volume = "110",
    number = "3",
    pages = "030001",
    year = "2024"
}

@article{Dokshitzer:1977sg,
    author = "Dokshitzer, Yuri L.",
    title = "{Calculation of the Structure Functions for Deep Inelastic Scattering and e+ e- Annihilation by Perturbation Theory in Quantum Chromodynamics.}",
    journal = "Sov. Phys. JETP",
    volume = "46",
    pages = "641--653",
    year = "1977"
}

@article{Gribov:1972ri,
    author = "Gribov, V. N. and Lipatov, L. N.",
    title = "{Deep inelastic e p scattering in perturbation theory}",
    reportNumber = "IPTI-381-71",
    journal = "Sov. J. Nucl. Phys.",
    volume = "15",
    pages = "438--450",
    year = "1972"
}

@article{Altarelli:1977zs,
    author = "Altarelli, Guido and Parisi, G.",
    title = "{Asymptotic Freedom in Parton Language}",
    reportNumber = "LPTENS-77-6",
    doi = "10.1016/0550-3213(77)90384-4",
    journal = "Nucl. Phys. B",
    volume = "126",
    pages = "298--318",
    year = "1977"
}

@article{Salam:2008qg,
    author = "Salam, Gavin P. and Rojo, Juan",
    title = "{A Higher Order Perturbative Parton Evolution Toolkit (HOPPET)}",
    eprint = "0804.3755",
    archivePrefix = "arXiv",
    primaryClass = "hep-ph",
    doi = "10.1016/j.cpc.2008.08.010",
    journal = "Comput. Phys. Commun.",
    volume = "180",
    pages = "120--156",
    year = "2009"
}

@article{deTeramond:2018ecg,
  author        = {de Teramond, Guy F. and Liu, Tianbo and Sufian, Raza Sabbir and Dosch, Hans G\"unter and Brodsky, Stanley J. and Deur, Alexandre},
  collaboration = {HLFHS},
  title         = {{Universality of Generalized Parton Distributions in Light-Front Holographic QCD}},
  eprint        = {1801.09154},
  archiveprefix = {arXiv},
  primaryclass  = {hep-ph},
  reportnumber  = {JLAB-THY-18-2630, SLAC-PUB-17217},
  doi           = {10.1103/PhysRevLett.120.182001},
  journal       = {Phys. Rev. Lett.},
  volume        = {120},
  number        = {18},
  pages         = {182001},
  year          = {2018}
}

@article{Buckley:2014ana,
	author = {Buckley, Andy and Ferrando, James and Lloyd, Stephen and Nordstr{\"o}m, Karl and Page, Ben and R{\"u}fenacht, Martin and Sch{\"o}nherr, Marek and Watt, Graeme},
	title = "{LHAPDF6: parton density access in the LHC precision era}",
	eprint = "1412.7420",
	archivePrefix = "arXiv",
	primaryClass = "hep-ph",
	reportNumber = "GLAS-PPE-2014-05, MCNET-14-29, IPPP-14-111, DCPT-14-222",
	doi = "10.1140/epjc/s10052-015-3318-8",
	journal = "Eur. Phys. J. C",
	volume = "75",
	pages = "132",
	year = "2015"
}

@article{dePaula:2022pcb,
    author = "de Paula, W. and Ydrefors, E. and Nogueira Alvarenga, J. H. and Frederico, T. and Salm\`e, G.",
    title = "{Parton distribution function in a pion with Minkowskian dynamics}",
    eprint = "2203.07106",
    archivePrefix = "arXiv",
    primaryClass = "hep-ph",
    doi = "10.1103/PhysRevD.105.L071505",
    journal = "Phys. Rev. D",
    volume = "105",
    number = "7",
    pages = "L071505",
    year = "2022"
}

@article{Karmanov:2009bhn,
    author = "Karmanov, V. A. and Maris, P.",
    title = "{Manifestation of three-body forces in three-body Bethe-Salpeter and light-front equations}",
    eprint = "0811.1100",
    archivePrefix = "arXiv",
    primaryClass = "hep-ph",
    doi = "10.1007/s00601-009-0054-3",
    journal = "Few Body Syst.",
    volume = "46",
    pages = "95--113",
    year = "2009"
}

@article{Ydrefors:2017nnc,
    author = "Ydrefors, E. and Alvarenga Nogueira, J.~H. and Gigante, V. and Frederico, T. and Karmanov, V.~A.",
    title = "{Three-body bound states with zero-range interaction in the Bethe\textendash{}Salpeter approach}",
    eprint = "1703.07981",
    archivePrefix = "arXiv",
    primaryClass = "nucl-th",
    doi = "10.1016/j.physletb.2017.04.035",
    journal = "Phys.~Lett.~B",
    volume = "770",
    pages = "131--137",
    year = "2017"
}

@article{Ydrefors:2022bhq,
    author = "Ydrefors, Emanuel and Frederico, Tobias",
    title = "{Proton quark distributions from a light-front Faddeev-Bethe-Salpeter approach}",
    eprint = "2211.10959",
    archivePrefix = "arXiv",
    primaryClass = "hep-ph",
    doi = "10.1016/j.physletb.2023.137732",
    journal = "Phys. Lett. B",
    volume = "838",
    pages = "137732",
    year = "2023"
}

@article{Mondal:2019jdg,
    author = "Mondal, Chandan and Xu, Siqi and Lan, Jiangshan and Zhao, Xingbo and Li, Yang and Chakrabarti, Dipankar and Vary, James P.",
    title = "{Proton structure from a light-front Hamiltonian}",
    eprint = "1911.10913",
    archivePrefix = "arXiv",
    primaryClass = "hep-ph",
    doi = "10.1103/PhysRevD.102.016008",
    journal = "Phys. Rev. D",
    volume = "102",
    number = "1",
    pages = "016008",
    year = "2020"
}

@article{Xu:2021wwj,
    author = "Xu, Siqi and Mondal, Chandan and Lan, Jiangshan and Zhao, Xingbo and Li, Yang and Vary, James P.",
    collaboration = "BLFQ",
    title = "{Nucleon structure from basis light-front quantization}",
    eprint = "2108.03909",
    archivePrefix = "arXiv",
    primaryClass = "hep-ph",
    doi = "10.1103/PhysRevD.104.094036",
    journal = "Phys. Rev. D",
    volume = "104",
    number = "9",
    pages = "094036",
    year = "2021"
}

@article{Xu:2022yxb,
    author = "Xu, Siqi and Mondal, Chandan and Zhao, Xingbo and Li, Yang and Vary, James P.",
    collaboration = "BLFQ",
    title = "{Quark and gluon spin and orbital angular momentum in the proton}",
    eprint = "2209.08584",
    archivePrefix = "arXiv",
    primaryClass = "hep-ph",
    doi = "10.1103/PhysRevD.108.094002",
    journal = "Phys. Rev. D",
    volume = "108",
    number = "9",
    pages = "094002",
    year = "2023"
}

@article{Lan:2019vui,
    author = "Lan, Jiangshan and Mondal, Chandan and Jia, Shaoyang and Zhao, Xingbo and Vary, James P.",
    title = "{Parton Distribution Functions from a Light Front Hamiltonian and QCD Evolution for Light Mesons}",
    eprint = "1901.11430",
    archivePrefix = "arXiv",
    primaryClass = "nucl-th",
    doi = "10.1103/PhysRevLett.122.172001",
    journal = "Phys. Rev. Lett.",
    volume = "122",
    number = "17",
    pages = "172001",
    year = "2019"
}

@article{Lan:2021wok,
    author = "Lan, Jiangshan and Fu, Kaiyu and Mondal, Chandan and Zhao, Xingbo and Vary, james P.",
    collaboration = "BLFQ",
    title = "{Light mesons with one dynamical gluon on the light front}",
    eprint = "2106.04954",
    archivePrefix = "arXiv",
    primaryClass = "hep-ph",
    doi = "10.1016/j.physletb.2022.136890",
    journal = "Phys. Lett. B",
    volume = "825",
    pages = "136890",
    year = "2022"
}

@article{Xu:2024sjt,
    author = "Xu, Siqi and Liu, Yiping and Mondal, Chandan and Lan, Jiangshan and Zhao, Xingbo and Li, Yang and Vary, James P.",
    collaboration = "BLFQ",
    title = "{Towards a first principles light-front Hamiltonian for the nucleon}",
    eprint = "2408.11298",
    archivePrefix = "arXiv",
    primaryClass = "hep-ph",
    doi = "10.1016/j.physletb.2025.139599",
    journal = "Phys. Lett. B",
    volume = "867",
    pages = "139599",
    year = "2025"
}

@article{Miller:2019ysh,
    author = "Miller, Gerald A. and Brodsky, Stanley J.",
    title = "{Frame-independent spatial coordinate $\tilde{z}$: Implications for light-front wave functions, deep inelastic scattering, light-front holography, and lattice QCD calculations}",
    eprint = "1912.08911",
    archivePrefix = "arXiv",
    primaryClass = "hep-ph",
    reportNumber = "NT@UW-19-20",
    doi = "10.1103/PhysRevC.102.022201",
    journal = "Phys. Rev. C",
    volume = "102",
    number = "2",
    pages = "022201",
    year = "2020"
}

@article{Collins:1981uw,
    author = "Collins, John C. and Soper, Davison E.",
    title = "{Parton Distribution and Decay Functions}",
    reportNumber = "OITS-162",
    doi = "10.1016/0550-3213(82)90021-9",
    journal = "Nucl. Phys. B",
    volume = "194",
    pages = "445--492",
    year = "1982"
}

@article{Broniowski:2007si,
    author = "Broniowski, Wojciech and Ruiz Arriola, Enrique and Golec-Biernat, Krzysztof",
    title = "{Generalized parton distributions of the pion in chiral quark models and their QCD evolution}",
    eprint = "0712.1012",
    archivePrefix = "arXiv",
    primaryClass = "hep-ph",
    doi = "10.1103/PhysRevD.77.034023",
    journal = "Phys. Rev. D",
    volume = "77",
    pages = "034023",
    year = "2008"
}

@article{Ma:2021tHooft,
    author = "Ma, Bailing and Ji, Chueng-Ryong",
    title = "{Interpolating 't Hooft model between instant and front forms}",
    eprint = "2105.09388",
    archivePrefix = "arXiv",
    primaryClass = "hep-ph",
    doi = "10.1103/PhysRevD.104.036004",
    journal = "Phys. Rev. D",
    volume = "104",
    number = "3",
    pages = "036004",
    year = "2021"
}
\end{document}